# Limits of Modal Insensitivity for Laser Vibrometry,
## Spectral Reduction Requires Super-symmetry

Michael C. Kobold[1]

**AUGUST 2014**

[1] Personal: matadorsalsa@startmail.com, Work: Michael.C.Kobold @ Navy.mil, Code X-12, Science, Technology, Analysis and Simulation Department, NSWC Panama City Division, Naval Surface Warfare Center PCD, 110 Vernon Ave, BL 110, 3B24W, Panama City, FL, USA 32407-7001, O 850.230.7212, (after-hours M 321.591.3679)

ABSTRACT

Laboratory measurements showed that identification (ID) and monitoring of objects using remote sensing of their vibration signatures are limited in a couple of rare cases. This work provides two necessary conditions to infer that the remote identification of common targets to within prescribed bounds is practical; failure to ID the spectrum is shown to be rare. Modal modulation of laser return produces data clusters for adequate spectral ID using slowly swept sine (SSS) and small deflection multi-modal (MM) analyses. Results using these completely different calculations lead to practical removal of a remote sensing concern, spectral "reduction" (SR) of return used for object ID. The optical return adequately provides structural mode ID for non-imaging detection and classification of vehicles. Calculations using a large spot size to completely paint the vibrating object provide insight for SR found in laboratory measurements which use spot size as a variable. Non-imaging analytical calculations comparing SSS and MM approximations match to second order. Numerical SSS and MM calculations show vibrating rectangular plates have spatially integrated (non-imaging) return that varies substantially among low-frequency vibration modes. The clustering of data from these two methods is a necessary condition for ID. The theories for SSS and MM describe the signal processing physics for a modal recognition capability and how pure one-dimensional modal targets and super-symmetric square plates can frustrate classification.

# Contents

## List of Figures

## List of Tables

**FOREWORD**

This extension of the author's Air Force Institute of Technology 2006 thesis provides mitigation of a concern in the laser vibrometry field. This stand-alone product is not a deliverable of existing projects but is a foundation for many. Variations of this type of sensing include interferometry sensing for buried object detection. This type of remote sensing is used and proposed by X-12 primarily for vehicle identification including a ground-based type of Identify Friend or Foe (IFF) to provide added safety to our troops during complex operations. Michael C. Kobold, PE, author.

**PREFACE**

The author would like to thank the US Naval Surface Warfare Center Panama City Division (NSWC PCD) scientists. For substantial review assistance: Dr. Paul Schippnick (retired) and especially Dr. Jeff Rish III. For beneficial discussions: NSWC PCD managers of ONR projects Dr. Harold (Dick) Suiter and John Holloway, University of Dayton Professor Partha Banerjee and Director Joseph Haus, Air Force Research Laboratory (AFRL), and Air Force Institute of Technology (AFIT) professors Stephen Cain (my thesis advisor), William Baker, Richard Cobb, and Robert Canfield. The author's thesis was encouraged by Dr. Matt Dierking (AFRL/SNJM), Dr. Robert Williams (AFRL/SNAT), and GD-AIS, whose support is greatly appreciated. Photos of the experimental apparatus, Figure 1 and Figure 2, were graciously authorized by Royal Australian Air Force (RAAF) Flight Lieutenant Ngoya Pepela whose AFIT thesis provides a pertinent experimental physical example of the insensitivity subject, spectral elimination.

**1.0 Introduction**

Laser vibrometry is a mature field. Definition of vibration modes is a purpose and region of engineering unto itself. Applications of methods to identify modes of vibration of distant objects continue to expand into different domains from microscopic to astronomical, from geological to geospatial, and from biomedical diagnosis to 'health monitoring' of potentially dangerous mechanical equipment. While ordinary acousto-optic modulation is in-surface aligned,[1] the use of lasers to probe vibration states (laser vibrometry) involves out-of-plane to near-normal reflection from the surface. Probe distances vary from standoff range (mm or cm) to remote sensing range (2 km $< R \lesssim$ 10 km). Astronomical radio vibrometry could assume an entire target (body of fluid) is illuminated. However, this article restricts analysis to the range of commercial laser vibrometers. This article shows that SR and SE problems can be mitigated.

Investigations of the quality of spectral estimation of a remote object's vibration state have uncovered a few issues. One of these issues is 'spectral *elimination*' (SE), a special case of the ordinarily less severe form 'spectral *reduction*' (SR).[2] For the purpose of this article use the laser vibrometry definitions; assume SR describes corrupted modal response while SE describes a condition of modal response where one or more of the modes is entirely missing.

The result of this paper validates that laboratory measurements with spectral reduction or elimination of modes for these *slender bars* or stiff strips (Figure 1) are repeatable. However, *do SR and SE for bars present a problem for adequate vibration spectral ID of other structures?* The conclusion of this article is that for structures that can be manufactured, SR and reduced spectral ID are insignificant or non-existent. To show this it will be necessary to fully describe SR, SE, and spectral uniformity (SU). These concepts appear in this article in several ways related to the types of analysis listed in Table 1.

This analysis shows that deleterious effects of this type of SR are limited to simple laboratory test samples or academically super-symmetric structures, as defined herein. The percentage of similarly symmetric structures that might be vibrometry targets can be considered insignificant. These potentially difficult targets comprise less than 0.01% of the complete set of typical target objects such as vehicles, windows, and the surfaces of bodies of liquids. In addition, while reports of spectral reduction in vibrometry[2] might seem to infer vibration-related object identification (ID) features are potentially untrustworthy (this has been a spoken fear in the industry), prior research[3] indicates that **consideration of multiple modes for classification combined with the variation in modal response shown herein help remove this concern.**

Super-symmetry is a condition seen in the set of vibration modes for perfectly square plates. In theory super-symmetry is easier to produce in vibrating bars with one-dimensional pure symmetry whose vibrations result in other symmetries *to be defined in the calculations*. Results in this report show that *only* super-symmetric structures lack of modal response variation. For all but these most unlikely of natural objects, **laser vibrometry provides overall adequate ID for vehicles by clustering of spectral features.** These features include the center frequencies of the structures' normal modes but can also include the response magnitude of those modes, or at least ratios of them (eigenvectors), in order to increase probability of proper classification. Without variation in modal response, the ability to classify structures wanes.

This paper has two calculations starting in section 2.0 on page 15 which provide theoretical insight into how spectral reduction and resulting in missed classification. The evidence shows that these SR concerns are unlikely to negatively affect the identification of target vehicles. Corroborating evidence includes the one dimensional (1-D) lab measurements of SR,[2] the SR seen in simulation of return from M1A1 tank armor plate,[3] and a Redstone Arsenal laser vibrometry project that remotely measured vibration signatures of a half dozen military vehicles (five types, many aspects).[4] Table 1 compares the three main analyses for remote sensing using laser vibrometry. The last type of estimate of sensed vibrations that this table compares is the set of (analytical) calculations in this report. The calculations are meant to encompass the assumptions of the simulation of return from M1A1 armor, but can be applied to quite general conditions far outside the measurements and simulations.

Table 1. Remote spectral signatures using laser vibrometry regions of applicability

| Estimate type | Target | Range | Project |
|---|---|---|---|
| Measurement (lab) | 1-D controlled bar | 1-2 m | AFIT Thesis, WPAFB[2] |
| Measurements | 5 military vehicles | 0.5-3.2 km | NATO study at Redstone Arsenal[1] |
| Simulation | M1A1 armor plate | 4 km | AFIT thesis[3] |
| Analytical | Generic | Generic | This report herein |

To help make remote sensing more universally useful, an emphasis on probe beams with *large spot size* (described below). This emphasis on large spot size led to this investigation of signals derived from spatial integration of the entire return (*non-imaging*).[3] The laser spot diameter is the width $w$ where power drops to $1/e^2$ of the maximum. *Small spot size* systems can also use *non-imaging* detectors but require tracking and pointing systems to estimate and control where the pin-point beam lies. For an $a \times b$ target area 'small' is $w \ll \min(a,b)$. Such a pencil beam must scan the entire object to ensure adequate representation of the lowest frequency modes on the surface. Large spot size systems, $w^2 > ab/\pi$, are less expensive and less cumbersome but require more signal processing to account for optical and image processing effects that are a result of spatially integration of the entire image field of view (FOV). Of particular concern in this paper are spatial phase correlations, the relationships of the phase at all locations along each vibration mode (on the target), as well as the measurement effect of the superposition of these modes on a non-imaging signal representing the response of the entire surface painted by the probing laser beam. The modulation across a target affects the phase of near-nadir directed rays, all of which combine together at the non-imaging detector. Later in the section that considers clutter, a schematic of this basic arrangement appears in Figure 3.

Measurement of the surface vibrations of vehicle modes ordinarily uses Doppler and dual pulse laser vibrometry systems, both of which utilize small spot size beams. In using these systems, laser vibrometry return from metal strips or bars in laboratory experiments have shown reduction or elimination of spectral modes for target irradiance return as spot size grows.[2]

The jargon is not uniformly formal: Sometimes the more extreme term spectral elimination (SE) is used to characterize even mild SR. The mechanism postulated to cause SE for such Doppler[4,5] or dual pulse[6,7] systems is spatial averaging over regions with opposing phase modulation.[2]

Extension to continuous wave (CW) phase modulation to measure coherent displacement harmonics would be subject to similar spectral reduction. This insensitivity investigation assumes CW illumination (not including pulsed measurements). The relationship between SR and SE compared to decreased detection due to spectral uniformity (SU, introduced in this article) is the purpose of this report – it is primarily a comparison of Figure 9 and Figure 10.

Organization:

This lengthy introduction section, continuing below, encompasses 35% of this report in order to introduce the 'Prior Measurements' and 'Prior Simulation' sections that follow. They are necessary to support the thesis that *ordinary (commercially available) laser vibrometry of vehicles and machinery is not subject to SE, SU, or enough SR to degredate detection and classification performance*. Two calculations introduced on pages 26 and 31 are the main part of this report. The foundations that simulation and measurements provide are important, but they are foundational only to establish the application for which these two calculations are useful. It is *the calculations* that provide a *partially direct proof that SR, SE and SU are not remote sensing ID problems*. More importantly they provide insight into why and how SR develops.

The Appendices further consider that systems required to fully subdue modes (active suppression) would be prohibitively intricate, hence assume that active suppression is not practical. The main calculations in this report assume these control systems are not part of the target system, that coherence is adequate and that the cross-spectral covariance (CSC) diagonal is adequately near unity.

## 1.1 Prior Laboratory Measurements

A 2003 laser vibrometry thesis[5] using pulsed laser vibrometers (commercially available) reported repeatable methods to develop SE. For the particular 'bar' structure in Figure 1 small spot size irradiance spectral content undergoes spectral reduction. This is a form of insensitivity of irradiant return to particular vibration modes when the spot size increases (Figure 2, photo). If a mode is fully missing (in the plots within Figure 1 → Figure 2 for the 1kHz line), it has undergone spectral elimination, a subjective assessment which is easy to verify. Both Figures include test equipment photos and spectra with permission.[8]

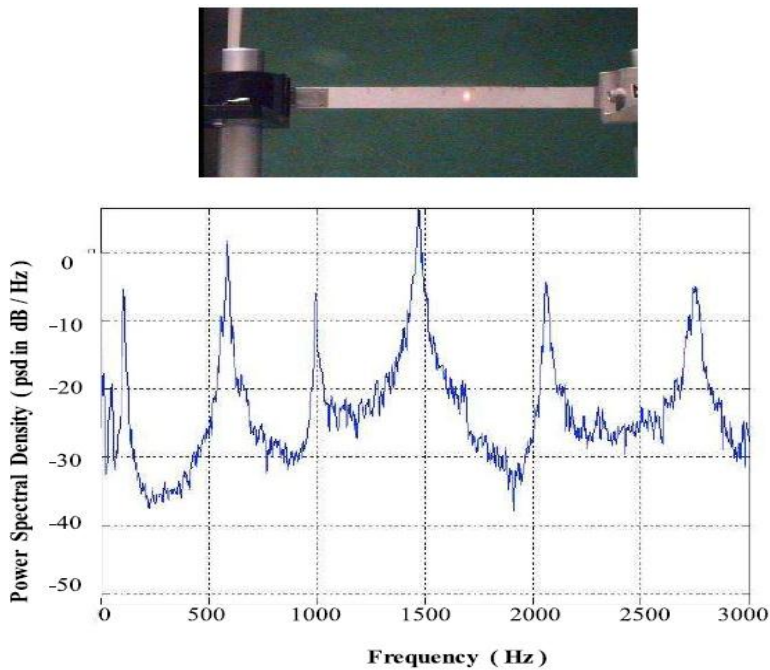

Figure 1. Reducing the spot size re-acquires the 1 kHz mode[2,8] (compare to Figure 2).

In the small spot size in Figure 1 above the 1 and 2.1 kHz modes are clearly active, but not in the larger spot size in Figure 2:

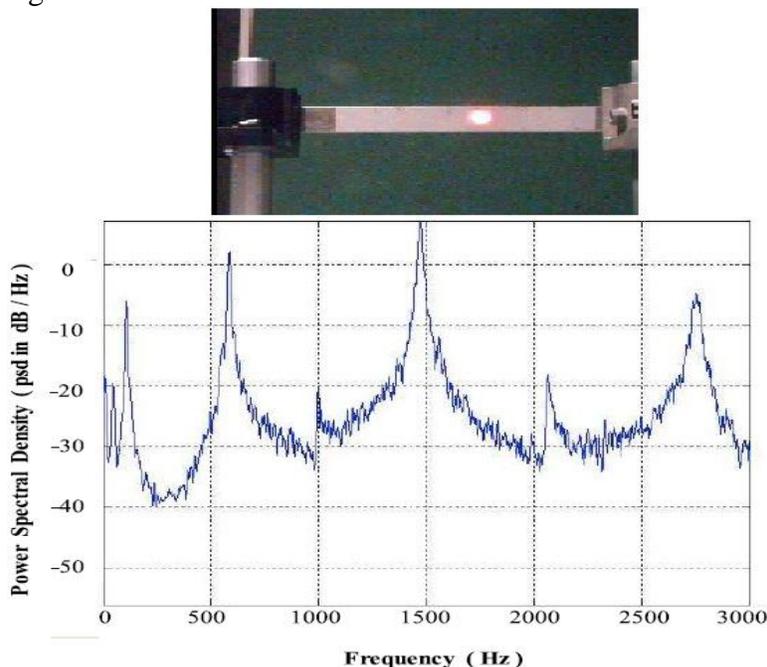

Figure 2. A large spot size removes the 1 kHz line[2,8] (compare to Figure 1)

Recreating this spectral insensitivity problem in structures that are not carefully crafted in a lab for this purpose is found to be impractically difficult because, for one, most targets are not likely to behave like bars in a laboratory. Apparently, only pure one-dimensional (1-D) and a finite few modes (defined in this article) within perfectly square 2-D vibration are capable of recreating this spectral reduction that produces deficiencies for laser vibrometry sensing. Assign the set of these two types of modes the name *super-symmetric* modes. These super-symmetric conditions are only practical to accomplish by computer simulation. The chance that a manufactured structure could attain such perfect symmetry becomes realizable only for the 1-D modes, the common modes of vibration in bars or strips of stiff material (herein called *bars*). Therefore, spectral classification of realizable structures does not have this deficiency.

Definitions of SR, SE, an SU – differences between different engineering communities

Spectral reduction and elimination (SR and SE) are terms that have expanded in use over the past decade into meanings that can differ. One example of a different form of SE is the use of notch band filters to reduce the spectral content of a modulation system.[9] Atmospheric seeing systems use similar spectral effects due to turbulence in the same manner as here. Bowman et al describe this form of SE: "the nonlinear coefficients are enhanced to account for the effect of the discarded modes on the explicitly evolved modes."[10] This is example where properly simulated turbulence requires a non-uniform modal distribution to the extent that modes are missing. Alternatively, the definition of SR and SE in this paper and in much of the laser vibrometry field describes an effect where remotely measured modulation differs from the actual target vibration due to the reduction or elimination of particular modes primarily due to an averaging effect seen as the spot size increases (Figure 1 → 2).[2] Figure 6 can infer this concept, but it is really meant to explain a more challenging but subtle issue in the foundations of modal analysis theory.

The types of ID and classification deficiencies listed in Table 2 below are inter-related by the environment in which they appear and the severity of the effect on probability of identification or proper classification. The table shows at which illumination the deficiency occurs. There is one case of Dependent Type (DT) that is another form of spectral uniformity, listed in column 2. The variables available as metrics that might characterize classification deficiencies include the area of the spot on the target, the modal bandwidth $BW = \Delta f = f_2 - f_1$, and the number of modes $N$. This table is meant as a partial road map for the different analyses that follow.

Table 2. Forms of spectral ID and classification deficiencies

| Type | DT | Description | Illumination | Variable |
|---|---|---|---|---|
| SE | | Spectral Elimination | Partial | Spot size |
| SR | | Spectral Reduction | Partial | Spot size |
| SU | | Spectral Uniformity | Full | Modal BW, $N$ |
| SU | USE | Uniform Spectral Elimination | Full | Modal BW, $N$ |

This report shows that SE and SR result from pure symmetries and SU from super-symmetries. SE and SR result in partial illumination of a target. Full illumination can result in spectral uniformity (SU) and uniform spectral elimination (USE) as a function of the number of active modes (bandwidth, BW). SU and USE result from a flawed form of frequency response function (FRF) calculation discussed later. A proper FRF without SR or USE can result in SU for very small deflection as the second calculation (MM) will show. SU is a form of FRF that is difficult to classify. Spectral ID for the purposes of this report relies on only one primary feature, the center frequency of each mode in a collection of modes. The classification deficiency occurs because the response value is uniform and thus provides no variation to cluster targets from non-targets. Non-imaging return is thus constrained to modal BW, as shown in Table 2. Different sections of this article will uncover different details of these interrelationships.

This analysis calculates the effect for a complete coverage of the target assuming the background spill-over clutter (Figure 3) is sufficiently far from the target or can be sufficiently time-gated or the 'spill clutter' effects can be otherwise removed. The Figure shows that the optical probe beam experiences a time delay between target return and clutter from 'spill-over' which might be exploited with time gating. Since the sensor gathers beam statistics in a reference ray shown at left, there may be other signal processing methods to reject large spot size spill-over clutter (*spill clutter*) by exploiting this *clutter delay*. However, such methods are not directly within the scope of this work except to assume that clutter removal was accomplished. Large spot sizes that fully paint the target for non-imaging sensing require less expensive equipment; there is no need for expensive fast laser pointing systems. We seek to determine if the full coverage illumination provides adequate target identification at this lower cost. Is there degradation of identification or detection power (probability of detection, $P_d$) at a given size (false alarm probability, $P_{fa}$)? Only target ID and classification sensitivity to SR, SE and SU are considered.

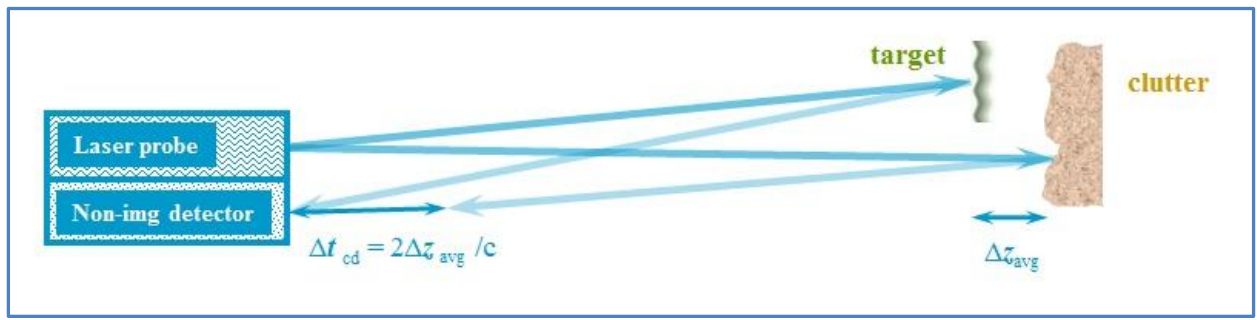

Figure 3. Between return from target and spill clutter is a time delay — filter clutter with time gating.

The fully illuminated target is assumed to be in front of the equivalent of a *green screen* so that the detection system processes only the modulated exitance from the target. Assume that some system, such as a time gating system using the clutter delay shown in Figure 1, already accomplished the clutter rejection so that the processing considered in the body of this report works only on the "green screen" data.

## 1.2 Prior Laser Vibrometry Measurements and Vehicle ID

Given the ease of identification capability for vehicles, the ID task should not be overly difficult. Use of only three modes at 0.1 Hz resolution can provide sufficient ID and detection power to classify five military vehicles of greater than $P_d \approx 95\%$.[3] Several studies[4] and personal experience with structural dynamics analysis for Ford, Fiat, Chrysler, and especially General Motors, establish that this form of multi-spectral recognition has been an intuitive part of engineering for more than half a century. With the advent of dynamic analysis required for various federal requirements mandating modal analysis, spectral identification of a vehicle make, model, and even year and age has been a common practice within production engineering since Modal Insensitivitythe late 1960's. With just a glance at a computer image of a simulation showing numerical values for the lowest four frequencies, a General Motors manager during a typical walk around, the truck engineering analysis group in Bloomfield Hills, Michigan was able to identify the 1993 Jimmy apart from several dozen models over a dozen model years that were being worked. Like fingerprints, low frequency vehicle spectral modes are that distinct.  From a view of only four values Mr. Bradshaw said to the author "So you are working on the '93 Jimmy."

The term *return* conveniently represents the surface modulated reflection, integrated over the entire detector (non-imaging) assuming the target was fully painted with a large spot size.  The term *radiant flux* represents the fully processed non-imaging signal, integrated over all pixels. This power signal in time is the basis for calculation of spectral response, an optical frequency response function (FRF).  The spectral response is sufficient for vehicle identification due to the unique 'fingerprint' nature of only a few low frequency modes in vehicle FRF's.[3,4]

The measurement systems used in the prior sub-section (Figure 1 and Figure 2) were HeNe 0.633 micron systems[5], whereas other related laser vibrometry systems used for the NATO study and another AFRL study operated at a wavelength of 2.019 microns.[4,6]

**1.3  Prior Simulation**

A simulation of an uncharacteristically symmetric structure validated the measurements listed in Table 2. Application to generic structures is not representative of the SR cases because in practice they would be less symmetric; even when coming off an assembly line within tolerance commercial structures are rarely optically flat and smooth with optically sharp edges.  Tolerances below a wavelength (optical wavelengths), sub-micron smoothness is impractical for commercial vehicles.  It is cumbersome to create structural variation of this microscopic nature in finite element (FE) models even though it would reduce compression stiffness and other effects by more than an order of magnitude.[40]  By default FE models have unrealistically uniform material, microscopically and optically straight edges and perfectly uniform thicknesses.  The effort to include the full variation at the microscopic level was limited to a hexagonal outer shape to model non-perfect cylindrical effects on the structural stiffnesses.

The structural Finite Element Analysis (FEA) is coupled to optical system. The optical model assumes the optical system is diffraction limited with a Strehl ratio of 100%. Aberrations and air turbulence are out of the scope of this work.  Further, the Fresnel diffracted return has a propagation[3] that assumes a 10 micron wavelength due to disk storage considerations. This arbitrary wavelength choice optimizes the target and detection grid densities to sufficiently characterize the target structural vibration frequency range within the capability of the computer systems in the AFIT Matlab processing cluster banks.  The 10 micron selection is an arbitrary data processing consideration that allows the entire propagation model to fit on a 4 GB RAM computer system.  However, from a physics consideration this system models wavelengths below 0.1 microns.  Since the structural wavelengths are so much larger than the optical wavelengths, changing the wavelength variable does not change the outcome qualitatively except to run out of space for smaller wavelengths.  The disk storage space requirements of processing a grid spacing sufficient to represent up to 500 Hz surface modes for the vibration simulation, and to propagate the modulation, leads to the need to arbitrarily model the laser wavelength as operating at 10 microns or more.

Time Fourier transforms of properly referenced total radiant flux time histories ($\Phi_e[t_i]$ in watts) provide the spectrum resulting from phase modulation of the laser irradiance due to the path length fluctuations between the laser system, vibrating surface and the detector system.[3]

The intermediate results of a simulation in Figure 4 show how different target deformation shapes (left) relate to different modulated return images (right).[3] The left pane of the Figure shows foreshortened deformed shapes from nonlinear contact FEA. The nonlinearities model wear, variation, and static indeterminacy statistics of the structural vibration. The simulation uses this deformed shape to modulate the phase of an assumed Gaussian probe beam whose exitance calculated via the paraxial D.E. and propagation via Fresnel diffraction provide the image signal received at the sensor. The right pane shows time-modulated optical $\Phi_e[t_i]$ images 4 km from the armor surface.

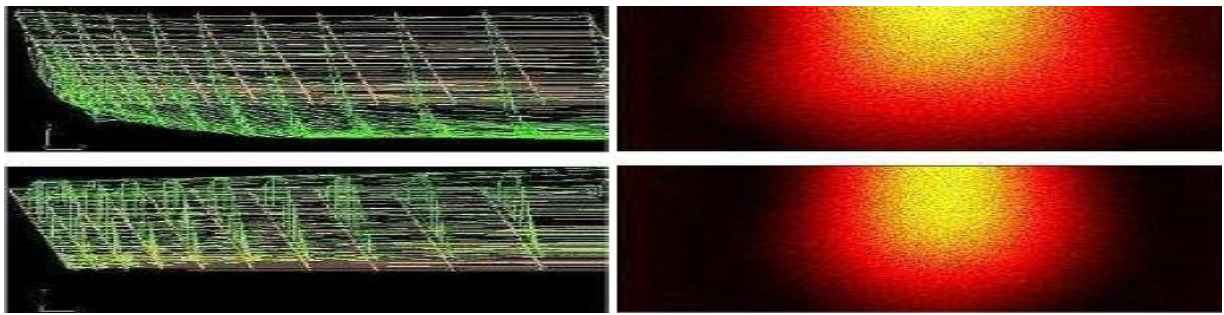

Figure 4. For two different times: at left are the FEA deformations, at right the received radiant flux

The left panes in Figure 4 display a view into the short (0.5 m) side of the mesh for a 1 m × 0.5 meter armor plate whereas the modulated propagated optical return to a 20 mm × 40 mm detector in the right panes is shown with the long 40 mm side horizontal; these images were oriented perpendicular to emphasize structural mode detail; otherwise the small FEA deformation is hard to see. One metric that provides an understanding of the coupling between remote sensing optics and the structural vibration is a cross-spectral covariance (CSC) composed of the mean-removed spectral densities $S_{MR}(f) = S_{MR}(f) - \mu_s(f)$ where $\mu_s = E[S(f)]$ is the mean value of the power spectral density (PSD) over variations of the structure. The ensemble averages (expectation values $E[\cdot]$) in this case were taken over variations in bolt fixity in order to simulate differences in age and tolerance stack-up. Frequencies $f_u$ and $f_v$ in equation (1) represent the PSD coupling between the structural ($u$ replaced by s) and optical ($v$ replaced by o) frequencies. Superscript T represents the transpose operation on the PSD vector while * represents the complex conjugate operation.

$$CSC_{uv}(f_u, f_v) = \frac{E\left[S_{MR}^T(f_u) S_{MR}^*(f_v)\right]}{\sqrt{E\left[S_{MR}^T(f_u) S_{MR}^*(f_u)\right] E\left[S_{MR}^T(f_v) S_{MR}^*(f_v)\right]}} \quad (1)$$

Figure 5 shows the diagonal of the CSC of the non-imaging optical response from the structural vibration of the same half-inch thick steel plate held by four corner bolts in nonlinear contact (open and closed surfaces) with a stiff base. For transfer function type representation used here the subscripts are the opposite to represent that the output is optical but here the first subscript is the abscissa found in the plots in the thesis,[3] kept as (s, o) and later changed to the transfer function form (o, s) after this CSC discussion. These CSC values are the normalized covariance between the structural vibration power spectrum and the modulated optical return at the sensor. The diagonal of the CSC, the variance for output frequency the same as input frequency, in Figure 5 has a mean = 0.999 near the ideal (unity), with a modest standard deviation of 0.673; the CSC does not vary sufficiently to degrade multi-mode classification. A fully linear response would be unity everywhere. This CSC plot shows that the optical non-imaging sensing of structural vibration for large spot size illumination is adequately close to unity so that spectral classification is not disrupted sufficiently to degrade classification by more than engineering error (5-10%) for all but the opportune case where the CSC spikes cover the important modal frequencies (see Appendix B). The simulation shown in Figure 4 resulted in a catalogue of results[3] that clearly show the modal features of the structure survive in the non-imaging return at the sensor for large spot probing of the M1A1 armor plate as a vehicle ID feature for IFF. Analysis of the suitability of this CSC for spectral ID appears in Appendix B on page B-1.

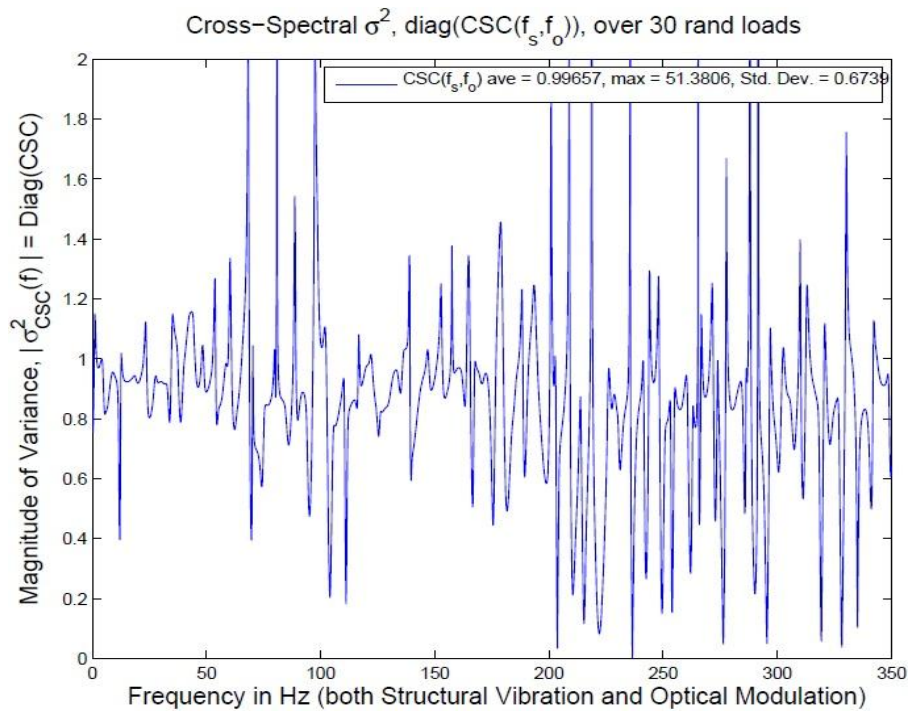

Figure 5. Diagonal of the structural-optical CSC: Spike outliers are few and narrow. $CSC_{avg} \approx 1$. This practical application of simulated remote classification shows that even the spatial averaging of the variation of the instantaneous spatial phase modulation provides an optical response spectrum $H(f)$ that is adequately close to unity to provide sufficient classification power within the lowest frequency modes of the vehicle transfer function for nonlinear vibration, which as previously described above allows for adequate vehicle identification. For linear systems the optical power spectrum $S_{opt}(f_o) = H(f_o, f_s) S_{vib}(f_s)$ in units of W/Hz is a product of the response $H(f)$ with the vibration power spectrum $S_{vib}(f_s)$ in units[11] of $g^2$/Hz when the signal $X_{opt}[t_i] = h[t_i] \otimes X_{vib}[t_i]$ is a convolution of the signal in time (discrete increments $[t_i]$) with the vibration in time $X_{vib}[t_i]$. Appendix B, the supporting thesis[3] and its references better define the CSC. To get an idea for the type of response the CSC describes, it helps to see it as a ratio. The numerator is a cross-spectrum $S_{o,s}(f_o, f_s)$ of $S_{opt}(f_o)$ with $S_{vib}(f_s)$, which is also a response of the nonlinear transfer function: $[S_{o,s}(f_o, f_s)]_{nonlinear} = H_{nonlinear}(f_o, f_s)S_{vib}(f_v)$. Linear cross-spectra provide insight, but are demonstrably in error. The cross-spectrum is strictly a time average over a period $T$ of the expectation value of the modulus of the product of both power spectral densities:[12]

$$S_{o,s}(f_o, f_s) \equiv \lim_{T \to \infty} \frac{E[S_{opt}(f_o)S^*_{vib}(f_s)]}{T} \quad (2)$$

While the optical modulation is not a linear function of the vibration power spectral density, the thesis shows how the spectra of the response shown at right in Figure 4 provide radiant flux whose spectrum is the nonlinear $S_{o,s}(f_o, f_s)$. This cross-spectrum is the result in the limit in Equation (2). Yet the nonlinear spectrum retains substantial features where their frequencies match those in the vibration spectrum $S_{vib}(f)$ even as the overall spectral response has a different shape. The source energy excites the linear normal modes with a nonlinear distribution of energy with mode frequency fluctuations that are usually negligible. Therefore the linear estimate of the optical output $[S_{opt}(f_o)]_{linear} = H_{linear}(f_o, f_s)S_{vib}(f_v)$ is useful as the linear time invariant estimate of the cross-spectrum. It is a matrix over the optical 'o' and structural 's' frequencies which becomes diagonal with the enforcement of $S_{o,s}$ linearity:

$$[S_{o,s}(f_o, f_s)]_{linear} = H_{linear}(f_o, f_s)S_{vib}(f_s)S_{vib}(f_s) \approx H_{Lin}(f_s, f_s)S^2_{vib}(f_s) \quad (3)$$

*This numerator can drive the CSC to zero; but the denominator going to zero can provide spikes.* In general the CSC normalizes this cross-spectrum with the square root of the product of the modulus of the two (|square|² = $SS^*$) square products $S^2_{opt}(f_o)S^2_{vib}(f_s)$ *as a denominator*. The simulation in the thesis combined structural nonlinearities in the FEA with brute force modulation providing power spectral densities $S_{o,s}(f_o, f_s)$ and $S_{opt}(f_o)$ that model nonlinear imaging. These optical PSD's combine with $S_{vib}(f_v)$ to provide a fully nonlinear estimate for the CSC. While the $CSC_{o,s}(f_o, f_s)$ diagonal plotted in Figure 5 is nearly always relatively close to unity, deviations from unity represent nonlinearities. The 0.7 standard deviation contains 51% of the spread of values within $CSC_{o,s}(f_o, f_o) = 0.7$ to 1.3 (using only $f_o$ for both frequencies which provides the variances, the diagonal of the CSC where $f_s = f_o$). However, it is the identification of modal *frequencies* that is important. The thesis[3] shows the nearly complete extent to which this nonlinearity of modulation does not impeded identification of vehicle modes; it shows that providing a CSC with 0.7 standard deviation provides lower than 15% false alarm and better than 80-85% probability of correct classification of moving and military vehicles,[13] std($CSC_{o,s}(f_o, f_o)$) = 0.7 is adequate and conservative. In this NATO laser vibrometer study these figures of merit represent *only* 3-5 frequencies used to classify the 5 military vehicles. Classification performance scales with the number of frequencies at a higher rate than a linear relationship would imply. (This would be a useful metric to measure in the future.)

Multispectral classification helps mitigate unlikely false alarm events where a narrow spike precisely covers a fundamental vehicle vibration mode. Figure 5 shows this type of false alarm appears to have a probability less than 0.1 %. False response is rare because ordinary structural modal response has local maxima that are much wider than even the bar modes in Figure 1 which have 25 × 500/17 ≈ 70 Hz wide responses. So the correlation with a much thinner spike is a relatively infrequent and small effect.

## 1.4 Adequate Modal Classification

From a classifier point of view, the CSC diagonal's deviations from unity (Figure 5) are similar to a partial corruption of fingerprints.[41] By eye and using branching maps prints are still easy to classify; fingerprints do not need to be perfect for useful assessment. Similarly, using multi-modal classification, laser vibrometry measurement also does not require such perfection.

Structural nonlinearities tend to help random excitation energy flow into deterministic vibration (normal modes), including plate contact modes.[14] Skin deflection varies at different modal frequencies for each mode in a superposition comprising the vibration time history. Nonlinearities affect modal participation factors (MPF's) and perturb resonant structural frequencies. However, deterministic and random loads excite distributions of modal frequency groups found in the linear normal modes, similar to how musical notes are composed of several different tones. The normal modes are eigenvectors that derive from the structural design. They respond in a sufficiently invariant manner[42] such that they are adequate system ID features — they form a stable basis set used extensively in automotive manufacturing and testing.[15,16,17,18]

These automotive and aerospace analysis methods use myriad load types ranging from monotone (sine sweep) through pseudo-random to excite linear normal modes. The assembly of active modes into a full response model incurs little MPF distortion. This MPF error has an effectively negligible effect on structural ID as was shown for impact and burst-random structural loads.[19]

The simulation thesis[3] studied the effects of propagation through atmosphere in the manner of a literature survey to provide more complete consideration. It is possible that large spot size sensing will average out issues with spatial coherence related to turbulence scale size within the 'pipe of air' between the sensor and target. While spatial coherence could be an issue for pencil beams used to sense 'high frequency' panel vibration above 300 Hz, the subject of this report, the thesis, and a recent NSWC PCD Technical Note,[20] is 'low frequency' vibration in the 50-250 Hz range where spatial coherence might not be an issue. A measurement would be useful.

The CW studies in this report used imaging simulation as well as non-imaging simulation[3] and measurement.[4] *These results suggest that compared to small spot size, use of large spot size laser vibrometry to sense low frequency vibration modes provides immediate, less expensive, less cumbersome application of laser vibrometry target classification algorithms.*

## 1.5 Modal Representation by Separation of Spectral Phase

Independent of time, the concept of modes of vibration is a difficult but essential consideration leading to a couple pedagogic formulas to help distinguish different forms of detection deficiency. To consider how closely-spaced the modes need to be for the equivalence of phase, the concept of standing waves would be convenient and is often valid. But an almost transient propagation in time of a single simple wave will be necessary. The math developed in the Optical Phase section indicated in equation (15), shows this concept of near independence. It helps to attempt a qualitatively conceptual understanding first, visually: In Figure 6 a probe beam, shown as a green swath, crosses over many cycles of a monochromatic mode (ripples on a surface beneath the beam swath) in one diagonal dimension $x$. Conceptually the mode is meant to appear to extend into the page as variation along the ordinate $x$ with an orthogonal time axis along that surface horizontally as the abscissa $t$. The beam is shown in uniform green, fanned out (a swath). The specular reflection (not shown) back to the sensor near the source introduces path differences, represented in the math as phase modulation by the vibrating surface. Phase differences are proportional to the ratio of the displacement at a point to the optical wavelength of the probe. The total return from the two different modes will have different modulation remainders (modulo $2\pi$). The integration over path differences is a function of fan length on the target. This length in $x$-$t$ space is a conceptual 1-D model for spot size, which in reality is an area in $x$-$y$ space. But that concept would require a three dimensional Figure.

Figure 6 is a spatial concept for 1-D modes in time. The location $x$ and time $t$ comprise a phase in the oscillatory representation where the electric field is $E(x,t) = A_o \exp[\theta]$ and the phase $\theta = j(\omega t + kx)$. In math symbols, represent the modal participation function MPF as $\varphi$. The Figure shows reality (left) versus small time increment $\Delta t$ approximation (right), holding the phase to small values ($\omega t \ll 1$) as will become apparent. A laser beam fans out (exaggerated) downward across different surface wavefronts for a given vibration mode (one eigenvector) $\exp \theta_k = \Sigma_i \exp[j(\omega_i t + \varphi_i k_i x)] \rightarrow \exp \theta_k' \equiv e^{j\omega t} \Sigma_i \exp[j\varphi_i k_i x]$. This approximation assumes $\exp[j\omega_a t] \approx \exp[j\omega_b t]$ for closely space modal radial frequencies $\omega_i = 2\pi f_i =$ a, b, ... so that time factors out of the sum in the phase term. This summary bears working out in detail. The separation of modes is a concept that is not usually proven in detail although it has been used by modal analysis engineers for centuries.[25] But it is sufficiently foreign to other engineers and scientists (until they see its use in their own fields) that this proof is apparently necessary to dissolve the resistance to analyzing nonlinear structures merely by their modal frequency fingerprints.

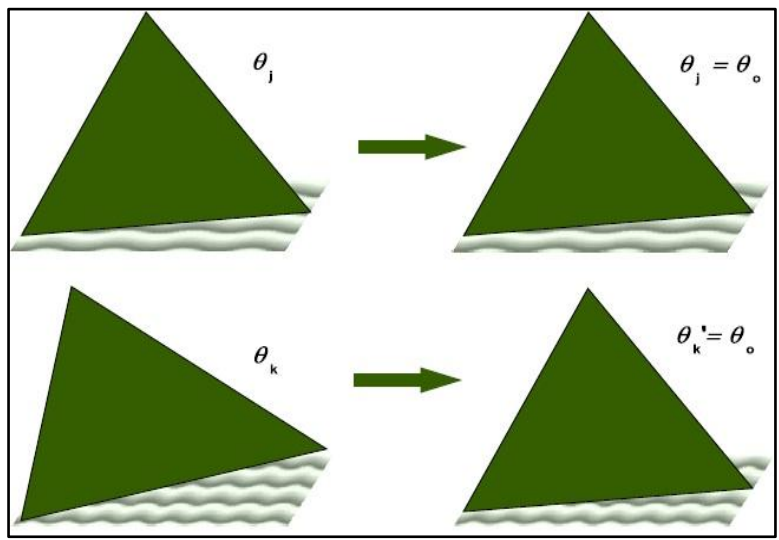

Figure 6. Condition for equivalent phase in two dimensions (wavy surface) illuminated from above

## 1.6 Time Independent Models, the subtle sorcery

How independent is time in laser vibrometry? Under what conditions can the spatial part of the phase separate from the temporal $e^{j\omega t}$ so that $\exp \theta_k \rightarrow \exp \theta_k'$, so that the upper row in Figure 6 is equivalent to the lower row within an adequate tolerance – so that rather than the coupling with phase and wave speed (tilt of the waves with respect to the wavy surface) we only need look at the wavelengths in $x$? Alternatively the time-based cyclic frequencies that generate those spatial frequencies $1/\lambda_i$ (or less formally $k_i = 2\pi/\lambda_i$) provide the equivalent modal basis set. Then the modal engineer after calculating the resolution (closeness of modes) can identify the structure based on the spatial wavelengths alone, as has been shown to be adequate in practice.[16] Assume modulation in space ($\exp[j\varphi_i k_i x]$) and time ($\exp[j\varphi_i \omega_i t]$) for a vibrating bar. For each plot in Figure 6 the abscissa (horizontal axis) has a uniform and very fine time phase but wavelengths along the ordinate vary. Looking downward the ordinate is the "vertical" axis, laid down by perspective diagonally fading into the background along the wavy surface in the four panes of the Figure. Later, within certain limits, (15) will show that for practical target modes this time modulation is approximately equal across several modes so that time ordinate has nearly the same scale portrayed on the right for $\theta_k = \varphi_k \omega_k t \rightarrow \theta_k' = \theta_j = \theta_o$. The slopes are the same and so we might characterize the underlying modulation by its spatial harmonics alone. Thus, separation of time and space is adequately small enough for modal analysis using the $k_i$ to characterize the modes. The calculation for the similarity $\exp[j\varphi\omega_a t] = \exp[j\varphi\omega_b t]$ in the Optical Phase section on page 25, (15) will arrive at this conclusion without any visual model such as Figure 6 being analyzed here.

The slope of the observation "cut" at the base of the green triangles models the temporal frequency of the individual mode. For clarity Figure 6 only represents two components of vibration with the low frequency mode on the upper row. Since the temporal phase is approximately equal in this case (discussed later for (15) on page 25)), the weighted space-time slope is approximately equal (all $\theta_j = \theta_o$) even though the wavelengths are unequal. This model assumes uniform time phase, approximately equal frequencies $\omega_{jk} \approx \omega_{j+1}$, even though mode locations in 'k-space' vary. As discussed later on page 25, if we choose a small integrating time $\Delta t$, the temporal phase between two modes can be made small ($[\varphi_j \omega_j - \varphi_k \omega_k]\Delta t \ll 1$).

Figure 7 introduces the uniform modal participation (UMP)[†] concept needed for modulation simplifications described later in more detail. This representation of UMP normalizes the amplitude, a common selection of eigenvalue result type from the typical options in FEA codes.[39] Other analytical developments use strain energy, but maximum displacement is easier to describe. It also simplifies this analysis to show one view of the system before considering non-uniform MPF systems. Combined with the UMP assumption (all $\varphi_i = \varphi_o$), the time variation of some modes drops out of the modulation equations for $2\pi\varphi_o \Delta t(f_i - f_{i+1}) \ll 1$ as discussed above in (15). The space coordinate separates similarly, which allows investigation of 'k-space' and '$\omega$-space' independently within the phase region just described. When dealing with only space modulation the analysis shows the combination of MPF and displacement amplitude is appropriate. This separation of dimensions ($x$ and $t$) provides one form of validity for the age-old mechanical engineering practice of modal engineers[19] to identify structures based only on structural wavelengths and mode shapes from test data represented by eigenvector plots in simulations.

---

[†] The details of Detection and Estimation theory have proofs based on a uniformly most powerful test[35,36] – but since we do not prove any detection theory in this paper, uniform modal participation should not provide an UMP acronym clash in this report.

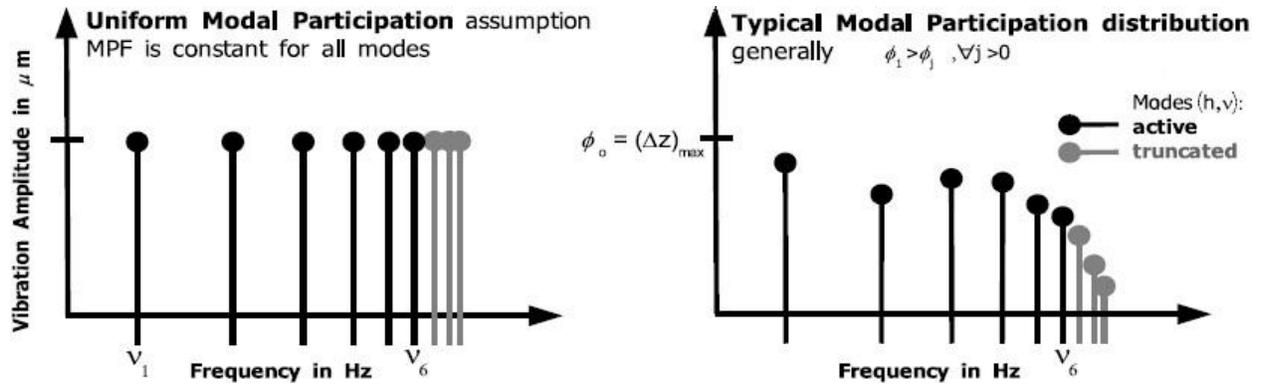

Figure 7. An UMP model (at left) helps analyze the mode distribution of a physical system (at right)

The frequency values shown in Figure 7 are the type of metric that has provided adequate classification even for UMP characterization (at left) of an example of reality (at right).[4,13] UMP uses all MPF $\varphi_k = (\Delta z)_{max} = \varphi_{o/N}$ a constant MPF over all $k$. The maximum deflection of the superposition of modes provides the approximated UMP modal deflection value:

$$\max_k [z_k(x, y)] = \max_k \left[ \sum_{k=1}^{N} \phi_k \right] = \frac{\phi_o}{N} \qquad (4)$$

This entire discussion and the need to prove the utility of modal analysis (using frequency values only) may seem elaborate. However, the mathematics used herein requires this detailed clarification of an age-old engineering practice, tabulation of modes without considering transfer functions, which are necessarily linear in application to most analysis methods like ABCD matrices of control theory and the extension[21] of convolution and correlation theories and relations related to Parseval's theory to frequency domain products of transfer functions with input signals. In the implicit use of the convolution theory $X_{opt}[t_i] = h[t_i] \otimes X_{vib}[t_i]$ in (3) to show that $[S_{o,s}(f_o, f_s)]_{Lin} = H_{Lin}(f_o, f_s)S_{vib}^2(f_v)$, a certain level of linearity was implied in order to factor out one $S_{vib}(f_v)$ from the optical response. Back on page 10 this was a careful construction while warning the reader of the existence of strong nonlinearities in the structural to optical system (subscript 'o,s'). The nonlinearities of the system must be applied at certain points in the theory of the response, while the application of linear concepts do in some cases provide some insight. The discussion must, as happens often in practice, use both the details and the survey view. Due to the multidisciplinary nature of this thesis, the experience of over a century of engineering practice, and the lack of cross pollination of these particular ideas across the optics and mechanical engineering domains, a detailed consideration of modal analysis is in order. In this case *modal analysis* exploits the utility of (usually) ignoring transfer function details in favor of estimation of the state of a system based on the frequencies that contain a minimum level of energy density above an arbitrary threshold, the *modes* of the system.

Consider the effect of spatial modulation across the profile of the CW probe beam on the return image at the sensor. Assume coherent imaging is spatially integrated into a non-imaging radiant flux in watts $\Phi_e[t_i] = \Sigma_{\Delta x, \Delta y}\{E_e[t_i][x_j\, ;\, y_k; t_i]$ watts/cm$^2\}$ piece by area $\Delta A = \Delta x \Delta y$ piece.[3] This can be a first step towards making a system less sensitive to airborne mounting vibration (without motion compensation). An additional reason to investigate CW vibrometry performance is that if spill-over clutter captured in the spot can be made benign (Figure 3), a large spot size that paints the entire target does not need as sophisticated a tracking system as current laser vibrometry systems using "pencil" beams require.[16] The rest of this report describes measurements, simulation and two approximations and discusses how they relate to calculated SR and SU based on UMP assumptions.

Simulation of laser beam parameter solutions using the paraxial wave equation model the Gaussian beam irradiance simulation.[3] A discussion of simulation results covers the source and statistics of the results. Following this discussion of the 'Simulation,' the subsequent section derives a plane wave solution and its assumptions. Finally an analysis of the two approximations, SSS and MM, over the low frequency modes shows where insensitivity resides: in the 1-D modes and an unlikely super-symmetry imposed on plates that are also perfectly square. These are the structures susceptible to spectral reduction. They are not representative of the typical targets for spectral ID.

## 2.0 Plane Wave Solutions

Assume a plane wave approximation for the optical irradiance incident on the vibrating target which has closely spaced structural eigenvalues ($\omega_i = \omega_{i+1}$). The upper ray of the schematic in Figure 3 models this plane wave propagation. This optical probe beam becomes modulated by a surface vibration due to path length differences by reflection from peaks and troughs of vibration, the structural standing waves.

For the structural vibrations of the target a propagation constant $k_k = 2\pi/\lambda_k$ with cyclic frequencies $f_k = c/\lambda_k$ of the eigenvalues $\omega_k^2 = (2\pi f_k)^2$ relates to vehicle surface vibrations at the fundamental mode ($k = 1$) and up to the cutoff mode ($k = M+N$). It is a well-known behavior[22,23] that the low frequency modes, $k = 1, \ldots \lesssim 20$ are well separated in the frequency domain below 100 Hz as sketched in Figure 7 to represent data such as that seen in the measurements[2,3,5,13] in Figure 1, Figure 2, and other references.[20,22,23,24] This low frequency separation contrasts with high frequency modes which vibrate at nearly indistinguishably closer frequencies; at the high frequencies end of vibration response most structural mode distributions have frequency maxima that tend to coalesce together into a continuous spectrum. The clustering sensitivity for spectral ID at discrete low frequencies observed over several centuries,[25] and of the returned radiant flux $\Phi_e[t_i]$ which is the result of the simulation provides an ability to classify different structures, as shown in the measurements.[4]

The lowest structural mode frequencies of these closed form relations are typically spaced closely enough for typical frequency response analysis.[4,16] The discussion of the closely spaced mode (CSM) approximation appears later on page 24. Under these conditions it is appropriate to use two useful theoretical closed-form methods. Their results turn out to model aspects of behavior equivalent to the AFIT experimental spectral elimination results in the limit of full illumination. These are the slowly swept sine (SSS) and multi-mode (MM) methods. The derivation of SSS-based plane wave modulation that follows produces Bessel function solutions. The Slowly Swept Sine theory and experimental results are comparable to a similar modulation using laser modulation in a birefringent crystal producing similar Bessel function solutions.[26] The MM method uses the small deflection assumption in order to obtain useful analytical expressions.

## 2.1 Geometry of the Optics Used to Classify Vibration Modes

Assume the detector and probing laser are at the same location so that vectors $r_1$ and $r_2$ are nearly parallel vectors which extend to appropriate points of maximum reflected phase difference (crest and trough) on the vibrating target object. Later in this section the constraint $M = 1$ models a rare 2-D rectangular 'super-symmetry' which allows spectral reduction. But for the moment assume that a 1-D vibration shape like that in part (A) of Figure 8 is the model under consideration. The number of modes along the $y$ direction is unlimited but we constrain the $x$ direction modes ($M = 0$) in order to represent the lack of modulation across the width of the strip or bar exhibiting 1-D motion. Setting the maximum mode $m$ to the limit $M = 0$ defines lack of vibration aligned with the x-axis, $z(x)$ = constant. But for now assume $M = 0$. This assumption excludes the large $M$ cases where high frequency modes $m > 0$ across a beam/bar might be excited because the bar is assumed to be short in the $x$ direction; these stiffer modes being ignored take more energy for the same deflection and even for plates "high energy" modes which have low strain energy[‡] are assumed to be in the set of modes that are cut off above the $M$, $N$ cutoff.

The potential use of laser vibrometry of interest to this remote sensing application is to identify (ID) vehicle type by sensing skin vibration. Structural mode shapes of vibrating vehicle skin are similar to the mode shapes of flat panels held with varied fixity on all edges. A suitable basic model used for the referenced simulation (Figure 4),[3] is a rectangular panel vibration that produces a 2-D phase modulation which is rectilinear due to the rectilinear boundary edges. An example of this mode shape geometry appears in the sketch in Figure 8. This is a target that is the most highly symmetrical target that could be used for vehicle ID, and impractically so; such structures are rare in commercial and military use. For plate structures with more than zero fixity on all edges, mode shapes no longer vary in just one dimension (Figure 8B).

---

[‡] These modes would take a lot of energy for the same maximum deflection amplitude, however, since there is little participation in such high frequency modes there is hardly ever much energy in them; they are low energy modes in the spectrum although a physicist might consider them "high energy" by physicists modes due to how the energy of a phonon is proportional to the cyclic time frequency.

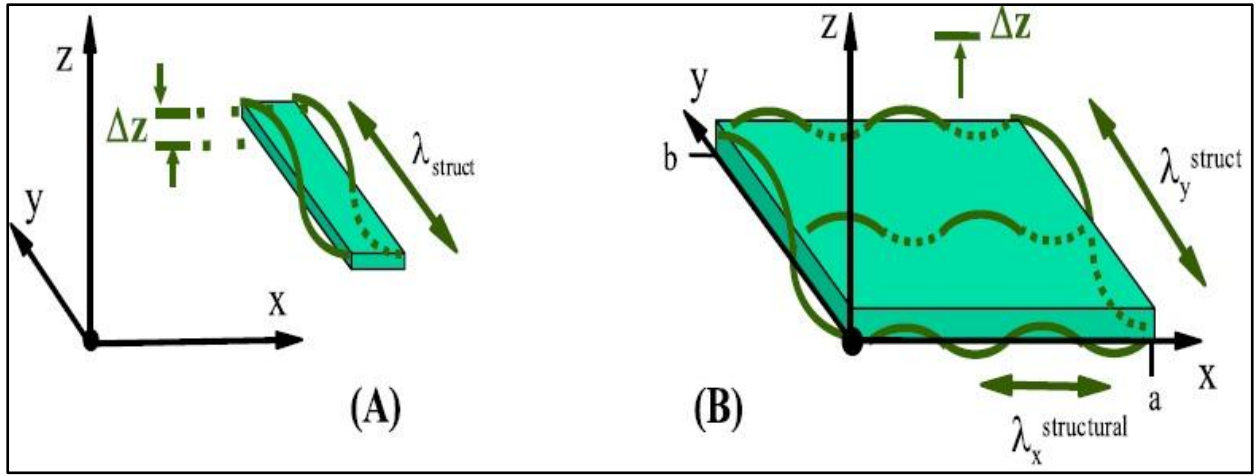

Figure 8. Modes shapes for (A) 1-D bar vibration $M=0$, $N=1$ and (B) 2-D plate vibration $M=2$, $N=1$

At the end of the first calculation the result of modulation from Figure 8 pane (A), a one-dimensional vibration mode shapes which is the $M = 0$, $N = 1$ case at left, is shown to produce an approximate result which is a Bessel function that is constant over all modes for the SSS model of phase-modulated spatially integrated return. In contrast for the 2-D system in pane (B), the non-imaging return modulated by two-dimensional mode shape $M = 2$, $N = 1$ cases at right results in variations of total radiance. In order to show the effect on the probe beam, a mathematical form of the deformed shape in terms of its phase change in the returned exitance will be the starting point. Equation (5) provides an estimate of the returned radiant flux using an approximately continuous integration over the surface (d$x$, d$y$) of area $ab$ keeping the summation related only to modes and the phase generic leaving the details of propagation for later calculation. The generic phase change upon reflection $\theta_r$ will be dropped and the phase change due to differences in deflection, the distance $|r_2 - r_1|$, are kept inside a generic deflection phase change $\theta_{j,k}$. Geometry details appear later (14). Inside the phase term is the sum of the modes using modal superposition,[25] a linear system idea that approximately applies to slight nonlinearities and to these models with the proviso that the modal participation factors $\varphi_{j,k}$ will vary in time.

$$\Phi_e[t_i] = \int_0^a \int_0^b E_e(x, y; t_i) e^{j\theta_r} \exp\left[ j \sum_j^M \sum_k^N \phi_{j,k} \theta_{j,k} \right] dxdy \qquad (5)$$

To sense these vibration modes a probe beam assumed to have a Gaussian profile illuminates the entire object. Integration over two dimensions ($x$, $y$), or summation over segments [$x_i$, $y_i$], of this typical Gaussian laser irradiance pattern $\Phi_e[x_i, y_i; t_i]$ with respect to the phase modulation (spatial sinusoidal modulation) across the target area results in non-imaging (total) return flux $\Phi_e[t_i]$ at the detector in watts. Non-imaging $\Phi_e[t_i]$ based classification capability would be a preferable phenomenology since it allows use of fast inexpensive sensors. The full flux is the integration of irradiance $E_e$ (W/cm$^2$) over target area, where the phase is summed over all modes. Equation 1 summarizes this concept assuming the return is propagated back to the detector. The result of the propagation is the output of a Fourier transform equivalent to Fresnel Diffraction, introduced later in (9).

## 2.2 Geometry of the Return, Sum of *N* Chladni Zones, Area Held Constant

Figure 8A displays the boundary condition used here where max(*m*) = *M* = 0, max(*n*) = *N* = 1. For some perspective into this concept of the superposition of modes[25] and its effect on sensed return, consider that $M > 1$ is a reasonable assumption for automated target recognition (ATR), a high frequency mode condition that is easier to classify because more modes are active. Targets in these calculations are simple two-dimensional $a \times b$ rectangular metal plates with $N_x N_y$ modes defined by integers: $[M,N] = [1, 1]$ through $[a/\lambda_x, b/\lambda_y]$ shown in Figure 8B. These integers *a* and *b* and the wavelengths $\lambda_i$ are a function of the structural dimensions and mode number. In these non-imaging calculations the integration of spatially oscillating irradiance $E_e(x, y; t) = d^2\Phi_e/dxdy$ over the area d*x*d*y* assumes the total flux $\Phi_e[x_i, y_i; t_i]$ (in watts) is a sum of an irradiance $(E_e)_{j,k}[x_i, y_i; t_i]$. (Not to be confused with the similar symbol for electric field *E* used later where energy and power are proportional to $E^2$.)

The main results of this report are shown at the end in the Chladni zone contribution using slow sine sweep in Figure 9 and the multi-modal estimate in Figure 10. These results use calculations for the SSS and MM plots based on the return from a consolidated deformed shape where the modes are summed into a resultant deflection vector in 3-D. The return from each point of reflection over the deformed shape propagates back to the detector where it is spatially averaged to simulate non-imaging detection. Alternatively, collecting return from a set of separate modes will assist in the analytical work that follows but requires some care with aspects of Guido Fubini's theorem.[27] The condition where the return per mode (or summation of return from several single-mode structures all at the different pertinent modes) sums to the total irradiance is *uniform modal participation* (UMP). Later analysis will clarify this concept mathematically. In preparation for those later calculations, the following discussion uses a summation of return for each mode where the modulation shape between nulls are regions whose size decreases with increasing cyclic frequency (in Hz). The regions are called Chladni zones. The deformed shape itself is a linear superposition of these modes. Under UMP conditions the return will be shown to also be a sum of return from different modes. The following calculation provides the coefficients for those calculations.

Assume the energy density incident on the target is uniform. However, intra-Chladni zone (anti-node) integrated radiant flux, per spatial wavelength, decreases with the number of spatial half-wavelengths along the bar. As $\lambda_i$ decreases, $\int_0^{\lambda_x} \int_0^{\lambda_y} E_e(x,y)dxdy$ represents smaller sensing patches, partitions to be summed. Higher modal temporal frequency increases the number of "modal" Chladni-zone patches ($f_{time} \uparrow \rightarrow f_{spatial} \uparrow$). But flux per node $\Phi_{e,zone}(t) = \oint E_e dA_{Chladni}$ decreases ($f_{time} \uparrow \rightarrow \Phi_e[zone] \downarrow$). These counter-effects result in (6) for segment *q*.

$$[\Phi_e]_q = \int_{(\Delta x)_q} \int_{(\Delta y)_q} E_e(x,y)dxdy \qquad (6)$$

For the 1-D vibration modes there are *n* segments $q = 1,n$ for each mode. The segments *q* range up to the current mode number *n* while the modes *n* range up to the cutoff number *N*.

Calculations of instantaneous return at the detector are in two forms, *slowly swept sine* (SSS) and *multi-modal* (MM). For SSS excitation (*calculation* 1) the sum of different segmentations of the same intensity will result in the same time averaged $\Phi_e$ for each mode acting alone. For a very small deflection MM assumption (*calculation* 2), non-imaging $\Phi_e[t_i] = |E_e|^2 \times$Area attains an asymptotic value for all but the lowest structural vibration modes. The two calculations SSS and MM described in detail later roughly envelope measured and simulated spectral return. For one-dimensional structures they show remote laser vibrometry insensitivity for 'bar' structures. In order to account for phase and interference the representation will now change from irradiance $E_e$ in W/cm$^2$ to electric field $E$ in V/m where the received power flux (irradiance) is proportional to the modulated field squared, $E_e \propto |E_{mod}|^2$(const.).

## 2.3 Bias in Slowly Swept Sine Response (SSS)

For laser vibrometry, *slowly swept sine* response is a popular calibration metric that *imperfectly* estimates the response to actual vibration due to nonlinearities of the system transfer function.[16] Calculations of *electric field E* using SSS are different from those for fully modulated field,[3] $E_{mod} \neq E_{swept}^{sine}$ as shown later in (16). Those relations indicate that SSS frequency response calculations based on the $\Phi_e$ return will model some vibrometry conditions. However, they do not properly characterize the coupled form of the received optical return.[3] SSS results clearly show a form of modal response insensitivity; subsequent calculations will show how solutions for E form a constant argument in the Bessel function. Spectra developed with SSS are usually meant for use as calibration metrics. In contrast, most spectral estimates use a system response that is fully coupled (MM), rather than a summation of monotone responses (SSS). It is in this respect that responses due to random vibration are often preferred,[16] in spite of the difficulty random analysis can present in the attempt to attain a converged solution with sufficient resolution. The error in SSS compared to the real FRF is a form of numerical bias.

Standard modal test equipment uses the SSS method extensively.[§] Derivation of MM response using assumptions provided in the next section and relations in leading to (16) act to expand the SSS results to full modal response. But MM is limited to very small deflections. That discussion will more properly justify removal of time and cyclic frequency from these equations for intensity and field return from the target as a consequence of tightly spaced modal frequencies. SSS is a useful pedagogic step before multi-modal analysis. SSS and MM both result in effects which provide mathematical models for the phenomenology of spectral elimination.

## 2.4 Assumptions of the Plane Wave Solutions

In its ordinary diagnostic application the *linear* response via *slowly swept sine* (SSS) uses return modulated by structural transfer functions known to be nonlinear in general. The Boundary Conditions (BC's) for the bar (structural 'strip') produce laser vibrometry return $\Phi_e$ that measures an integer number of vibration half-cycles. There are $n+1$ anti-nodes ($n$ nodes) for any group of $n$ half-cycles in a given dimension, each of which is located between the Chladni lines of nodes[25] across the width of the 'bar.' For example, Figure 8A has 2×0 anti-nodes (two half cycles) whereas Figure 8B has 2×4 anti-nodes (3×5 nodes). Equation 2 shows that the more anti-nodes there are in a given length (the higher the structural vibration frequency) the smaller is the energy per segment (anti-node).

---

§  Brüel & Kjær, www.bksv.com and Vibration Research, www.vibrationresearch.com (links circa 2006)

For a Chladni zone[**] of area $A_{cz}$ within a plate, sized $a \times b$ of area $A_{max} = ab$, we sum over Chladni zones ($\Sigma_{\text{C zones}}$) *of all sizes* in (7) where $A_{\text{Chladni}}$(zone) is the area for a particular Chladni zone. All Chladni zones have anti-nodes at their centers. Since the optical return from the full edge-constrained physical model does not provide constant phase modulation, time modulation of the field and exitance occurs as described below and implied in the mode shapes in Figure 8. The integration of exitance $M_e \propto |E_{mod}|^2$(const.) over area becomes $\Phi_e$ which has units of watts. A radiometry note: after exitance $M_e$ leaves the target (exits) and approaches the detector it changes name to irradiance $E_e$ that irradiates the sensor. It is proportional to the electric field squared. In calculations which follow $E_{in}$ and $\tilde{E}_{max}$ represent field densities in V/m per area over a small patch of area $A(q) = (\Delta x)_q (\Delta y)_q$ so that they have units of (V/m)/cm$^2$ to be summed with other patches to calculate the return. In (7) the vector variable $E$ is the electric field falling on the detector.

$$|\vec{E}| = \frac{\tilde{E}_{max}}{A_{max}} \sum_{\text{Ch.zones}} A_{\text{Chladni}}(q) = \left( \sum_{Cz\, q=1}^{MN} \frac{A_{cz}}{A_{max}} \right) \tilde{E}_{max} \qquad E_{\text{Chladni}}(q) = \left( \frac{\frac{a}{M} \frac{b}{N}}{ab} \right) \tilde{E}_{max} = \frac{\tilde{E}_{max}}{MN} \qquad (7)$$

For a target with approximately uniform illumination the value of integrated radiant flux related to exitance from a 'cell' (anti-node, Chladni zone) on that surface, which has the length of half a vibration wavelength, will decrease with the number of spatial half-wavelengths along the beam. Radiant flux $\Phi_e(t)$ over a Chladni segment (3) becomes smaller for higher vibration mode frequencies. This anti-node region of area constricts x-direction mode shapes since $\lambda_x \times \lambda_y \rightarrow$ width$\times \lambda_y$ for the $M = 0$ by $N$ modes for a one-dimensional (1-D) bar vibration system; the short width drives the energy requirement for $M = 1$ modes too high for any excitation to bleed into those modes, they are cut off by modal participation, MPF$(M>0,N) = \phi_{M>0,N} \approx 0$.

This 1-D vibration system has $N$ modes along the $y$ direction with no flexure across the $x$ direction. The actual 1-D deflection function is a full linear combination of $N$ one-dimensional (1-D) modes along the y axis. Viewing each mode separately is standard practice in modal analysis. As long as the bar stays approximately flat, maintains small deflection and small angle assumptions ($\Delta z \ll \sqrt{A}$, $\theta_{\text{surface}} \approx \sin \theta_{\text{surface}}$), the SSS result is that the sum of the return from a number $M_x N_y = ab/\lambda_x \lambda_y = ab/A_{\text{Chlandi}}$ of these Bessel functions remains constant. This is the result that will evolve into (13) from the discussion that follows. For a vibrating bar or strip $M_x = 0 \leftrightarrow$ $\lambda_x = \infty$ since deflection along $x$, $w(x)$, is "DC" (indicating no oscillation, uniform deflection). Thus the number of 1-D modes $M_x N_y$ for one-dimensional vibration functions becomes:

$$M_x N_y \Big|_{M_x = 0} \equiv N_y = \frac{b}{\lambda_y} = \frac{ab}{A_{Chladni}} \quad (8)$$

---

[**] Lord Rayleigh[25] reports (page 347) that it was Poisson who first *successfully* considered free vibrations of membranes [Mém. De l'Adadémíe, t. VIII 1829] which apparently drew Kirchhoff's similar solution. The solution for circular membranes was published by Clebsch [*Theory of Elasticity*, 1862]. Savart postulated that a membrane can respond to any pitch (driven excitation) encountering an apparently opposing set of solutions from mademoiselle Bernard of The French Academy. Experimental studies by Elsas [*Nova Acta der Kcl. Leop. Carol. Deutschen Akademie*, Bd. XLV. Nr. 1. Halle, 1882] confirmed [some] conclusions of Savart (Rayleigh's page 349). But the concept of superposition deserves special consideration only a quote can provide: "Doctor Young and the brothers Weber appear to have had the idea of superposition as capable of giving rise to new varieties of vibration, but it is to Sir Charles Wheatstone [89] that we owe the first systematic application of it to the explanation of Chladni's **Figure**s." From Rayleigh[25] page 377.

For mode $(m, n)$ the Chladni zones of area $A_{cz}$ within a plate of area $A_{max} = ab$ there is an electric field $\tilde{E}_{max} = mn |E| A_{Cz}/A_{Chladni}$ where the $\tilde{E}$ is the sum over the electric field increments (or integration of densities) from all zones. The incident field $E_{in}$ specifies the amplitude *before* modulation, in units of volts/meter. For the two-dimensional plate vibration case (2-D) with $M$ half-modes along the $x$ axis, $N$ along $y$, (9) provides the modulated electric field $E_{mod}$ derived from segments of scattered or reflected field (density) $E_{max}$ *just before* back propagation to the detector. For any one mode, the incident field density is $\tilde{E}$ for each of the $M \times N$ anti-node Chladni zones (7). For a superposition of modes the coefficient of $E_{max}$ changes from $A_{Cz}/A_{max}$ to the integral of space over the sum of modes shown in (9). The phase term of (9) already assumes normal incidence discussed later in (14). $\varphi_r$ is a common reflection phase related to $\theta_r$ in (5).

$$\vec{E}\Big|_{M,N \neq 0} = \int_0^a \int_0^b \frac{\vec{k} \cdot \vec{r}}{kr} E_{in} \exp\left[-j \frac{4\pi}{\lambda_{opt}} \left( R + \sum_{m,n=1}^{M,N} (\Delta z)_{m,n} \sin \frac{\pi x}{a/m} \sin \frac{\pi y}{b/n} \right) + j\varphi_r \right] dx\, dy \quad (9)$$

Time modulation appears later on in this report; this section mostly discusses the effect of spatial mode shapes, the major concern in analyzing spectral reduction and uniformity (SR and SU). For convenience assume the reflection phase $\varphi_r$ zeros the overall phase at Chladni lines ($\sin\theta_x$ and/or $\sin\theta_y = 0$) to remove the negative '–' from the exponent of (9). Assume a particular instant in time and choose a field at range $R = |\vec{r}_{avg}| \rightarrow E'_{in} \exp[j4\pi R/\lambda] \approx E_{in}$ so that only the modulation about the average range remains in the phase term. Using out-of-plane deflection $w_{m,n}(x, y)$ which is the sum of all the active modes (the summation in (9)) the range part $R$ factors out:

$$e^{j\frac{4\pi}{\lambda}[R + 2w_{m,n}(x,y)]} \approx (\text{const.}) e^{j\frac{4\pi}{\lambda} 2 w_{m,n}(x,y)} \qquad w_{m,n} = \frac{1}{2} \sum_{m=1}^{M} \sum_{n=1}^{N} \Delta z_{m,n} \sin \frac{\pi x}{a/m} \sin \frac{\pi y}{b/m} \quad (10)$$

*This interferometric choice of R assumes the detector is stationary*. Otherwise the deflection variation can average out through zero-mean jitter. But motion compensation is not trivial. Interferometric measurements such as this can measure deflections on the order of a fraction of the optical wavelength $\lambda \equiv \lambda_{opt}$. Variations in range $\Delta R$ comparable to 5% of $\lambda$ (0.4 to 0.7 micron relates to the visible range from violet to red) will provide a measureable phase shift; mounting platform variation in time has been a problem in the past.[7]

The return modulated field $E_{mod}$ from a rectangular plate in (9) shows the general phase modulation as a sum of normal modes $w_{m,n}(x, y)$ defined in (10). A "uniform MPF assumption" (UMP) will equalize individual deflections $w_{m,n} \rightarrow \varphi_{m,n} (\Delta z)_{m,n}/2$ for each mode where all $\varphi_{m,n} \rightarrow \varphi_o$. The modal participation factors (MPF's, $\varphi_n$) are related to the deflection oscillation ranges $\max(\Delta z)_{m,n}$. Use of simplifying assumptions such as UMP can be a reasonable method to produce a practical result where none otherwise exists. UMP assumptions used in vibration modal analysis are similar to uniform "mode energy distributions" used in laser engineering[28] which allows for practical analysis of the physics.

An alternative form of (9) that hints at a simple closed form sine swept solution appears in (11) which factors out the modulation per mode. This is still an introduction to the idea for slowly swept sine (SSS) to be accomplished in the next section (SSS on page 26). For later integration into a Bessel function solution, *the product can only be over a single term.* The exponential of the deflection sum does not turn into a product of individual exponentials unless the modal participation is identical for each mode, point by ($x,y$) point. The reason SSS is not a true FRF begins to become apparent. Test laboratories use SSS measurements equivalent to these to assemble spectra due to monotone excitation, one frequency band at a time, as do the formulas[3] resulting in an assembled modal solution $E_{SSS}$ summarized later in the main SSS result (19) based upon this modulation field, $E_{mod}$.

$$E_{UMP} \approx \int_0^a \int_0^b \prod_{m,n=1}^{M,N} e^{-j\frac{4\pi}{\lambda_{opt}}\left(\frac{1}{2}(\Delta z)_{m,n} \sin\frac{\pi x}{a/m} \sin\frac{\pi y}{b/m}\right)} dxdy \qquad (11)$$

In either case, SSS or multi-mode (MM), the wave vector ($\mathbf{k} = 2\pi\hat{\mathbf{e}}_k/\lambda$) and distance vector ($\mathbf{r}$) used in $(\Delta z)_{m,n}$ are parallel enough so that $\mathbf{k}\cdot\mathbf{r} \approx kr$ (shown later in (8)).
For the remainder of these calculations, arbitrarily assume a field polarized parallel to the $x$-axis unit vector, $\hat{\mathbf{e}}_k$. For the sum of modes in the exponent of (9) to factor into a product of exponents seen in (11), there is a requirement: **SSS results assume a uniform modal participation** (UMP) over N modes in y and M modes in x, providing an approximate amplitude per mode of $w/(MN) = (\Delta z)/(2MN)$. Here $\varphi_o=1/MN$, which is the UMP model. Application of the simple Bessel function definition[29] or its half-cycle form[30] from (12) applied to (11) provides the return for one particular mode, $m = m_i$, $n = n_i$ in (13). Calculating the field in a single Chladni zone measuring one half spatial wavelength in each direction (with Chladni area $\lambda_x \times \lambda_y/4$) provides a foundation for the field incident on the entire plate and the resulting spatially integrated radiant flux returned to the non-imaging detector.

$$J_n(\xi) = \frac{i^{-n}}{\pi} \int_0^\pi e^{i\xi\cos\theta} \cos(n\theta) d\theta \qquad\qquad J_o(\xi) = \frac{1}{2\pi} \int_0^{2\pi} e^{i\xi\cos\theta} d\theta \qquad (12)$$

For any single mode of a 1-D vibration where only *y* vibration displacement is active (Figure 8A), a half cycle of the spatial wavelength representing one Chladni zone for that temporal frequency, the modulated field integrated over that area, the half-cycle $E_{cy/2} \equiv [E_{in}(m,n)]_{cycle/2}$ is a constant since integration of sinusoidal modulation (see (12)) is the kernel of Bessel functions.[30] Coefficients of the Bessel function in $\Phi_e$ arise from using (9) to represent field density for each segment resulting in the integrated field in (13) whose squared value is proportional to $\Phi_e$.

$$E_{cy/2} = mn\frac{E_{in}}{mn} J_0\left(-2\frac{2\pi}{\lambda}\frac{(\Delta z)_{max}}{2MN}\right) = E_{in} J_0\left(2\pi\frac{(\Delta z)_{max}}{2MN\lambda}\right) = \text{const.} \qquad (13)$$

The full SSS specular irradiance derived from the half-cycle field $E_{cy/2}$ simplifies because for $M=0$ the solution is a Bessel function with an argument that is constant as shown later in (17). *But first*, there is a need to tie down some structural standing wave details with a bit of Chladni zone algebra.

## 2.5 Structural Considerations of Integration Over Chladni Zones

Under small strain assumptions to allow for "linear" superposition of modes, consider the largest energy component of a non-UMP system, the mode with the most *modal participation* $\varphi_{max}$. (Not to be confused with the radiant flux in watts, $\Phi e$. See page C-1.) Variation of highest frequency content (non-negligible modal participation) along $x$ determines $M$, and similarly along $y$ for $N$. Assign $\lambda_{y.high}$ to be the highest low strain energy $y$-mode under consideration and similarly for $x$. Just to clarify, these wavelengths represented by lambda with $x$ and $y$ subscripts are on the order of a meter, whereas the optical wavelength, lambda with no subscript or 'opt,' is less than a micron. The rectangular vibrating plate with $M = 2a/\lambda_{x.high}$ half-modes in the $x$ direction, and $N = 2b/\lambda_{y.high}$ half-modes in $y$, has a fundamental mode $M = 0$, $N = 1$ for $\lambda_{x.high} < \lambda_{y.high}$ (and vice versa) unless opposing edges are not of free fixity. In that latter case the fundamental will usually be $M = N = 1$ for common structures. This is sometimes referred to as the main "diaphragm" or "drum-head" mode which has only one half cycle along each edge.

If either mode number is a "Direct Current" (DC) type mode, $M = 0$ or $N = 0$, then the equation for the returned field $E$ requires a slight modification: removal of the pertinent DC mode from consideration, and using the other axis as the only one along which modulation occurs. The difference appears later in the separation of $M = 1$, $N = 1$ solutions in (19) and (20) and from the simulation results[3] where the first mode is number 1 in the sum.

1-D modes ($M = 0$) do not have a solution format that directly reduces from the most restricted 2-D solution ($M, N > 0$). Hence, the unharmonious name, DC. They have no sum at all, similar to $M = 1$ in that there is one term, but it is un-modulated (reflection from a uniformly flat surface). To modify (9) for the optical return from vibrating strips with one-dimensional modes (free-free condition on opposing long sides), consider that the number of structural modes perpendicular to the length of the strip is zero. The Matlab code cannot apply $M = 0$ or $N = 0$ to the modulated field in (10) for removal of modulation in one "dimension" of the mode shapes. This is an algorithmic limitation and requires a different formula for DC modes (no modes in a particular dimension). The Bessel function solution in (13) is the $M = 0$ as shown in Figure 8A. It shows a "sine swept" formula for electric field modulation by a vibrating strip aligned parallel to the y axis with the short edge width along the x axis, where free–free unconstrained BC's are along the two long edges (planes $x = x_1$, $x = x_2$) as shown in Figure 8A.

The "DC case" M = 0 calculation of the phase modulation of the returned field excludes $M$. The edges aligned parallel with the y axis have free–free BC's so no modulation along $x$ is possible. Therefore, there is no summation over the x dimension for return from the metal strip. Simulation[3] and theoretical results in this paper model optical response from sine swept vibration (SSS) only as a modally uncoupled system. While OEM's avoid modal coupling in vehicles and components, the coupling that does occur in multi-modal structures can produce substantial measurement error in sine swept estimate of the radiant flux $\Phi_e$ (in watts).

So for the math, when $M = 0$ then $MN \to N$. The bar vibration $M = 0$ has a mode count $N$ because the number of modes $= N$, not $MN = 0 \cdot N$ but rather the count only uses the active dimension, $y$. So the calculation for return from bars (1-D vibration) cannot use the same formula for surface vibration (2-D). Nevertheless, for generality we use the mode count for full surface vibrations, $MN$, assuming both $M, N \neq 0$ unless otherwise specified.

In the uniform modal participation model the maximum mode number, $MN$ for plates or $N$ for bars, is an indicator of a form of modal *bandwidth* that identifies the response in a non-uniform discrete frequency space. For UMP vibration modes $r = 1, 2, 3, ...$ frequencies $f_r \neq r f_o$ in practice. The order of $r$ will map to the $n^{th}$ or $(m, n)^{th}$ mode which has that frequency $f_r$. Although there is a deterministic distribution of mode frequencies due to eigenvalue solutions of the mass-stiffness matrices, the locations of the frequencies are often assumed to be random. However, within the SSS derivation of non-imaging $\Phi_e$ there will be an assumption of a uniform linear distribution of sufficiently closely spaced modes, sufficient CSM. This is an UMP and CSM modal system.

## 2.6 Optical Phase, Time-independent Relations

The optical probe beam shining on the assumed specular surface undergoes a phase change equal to twice the deflection amplitude in wavelengths, not including the reflection phase change.[26] $\Delta \varphi = 2kw(x, y) = 4\pi w(x, y)/\lambda_{opt}$ is the optical phase change in radians produced by path differences crest-to-trough due to standing wave structural displacement along the profile plane range $z = r \cos\theta_{incident}$. Equation (14) describes this spatial phase relation where $r_1(x, y)$ and $r_2(x, y)$ are optical path distances to two points on the target. Set a zero deflection range between crest and trough arbitrarily assigned to $r_2 = r_2(x_2, y_2)$ as a *reference distance* for zero deflection amplitude, $w' = w(x_2, y_2) = 0$. For any one mode $\Delta z = 2w$ where $w(x, y)$ is the vertical plate deflection field[3] amplitude assuming a suitable reference deflection $z(r_1)$.

$$\Delta \phi_{\bar{r}_1, \bar{r}_2}(x, y) = 2\vec{k} \bullet \vec{r} \approx 2k(r_1 - r_2)\cos(\theta \approx 0) \approx 2\frac{2\pi}{\lambda} w(x, y) = \frac{2\pi}{\lambda} \Delta z(x, y)$$

$$r_2(x, y) = \bar{r} = \frac{1}{T} \oint_{0...T} r_2(x, y) dt$$

(14)

In the rest of the paper $\varphi$ is the common symbol for the modal participation factor (MPF) following vibration analysis and finite element analysis (FEA) conventions. In the derivation of spatially modulated electric field $\boldsymbol{E}_{mod}$ (9), the time modulation $\mathcal{R}e\ [e^{j2\pi f_n t}] = \cos \omega_n t$ factors out before selection of an instant in time.[3] Arbitrarily choose $t = 0$. Structural modes are not precisely spaced close enough in frequency space for this assumption to hold formally, but it is adequate for our purposes. Tabulated Pininfarina chassis frequencies[22,23] and Laser Doppler vehicle frequencies[4] show this discrete nature of low frequency modes. So time variation does not strictly factor out of the exponent. But for *closely spaced modes* (CSM) this error is small (9). Using typical suspension mode variation from 17 to 33 Hz, with second mode frequencies usually nearby at 25 to 55 Hz, the CSM assumption for automotive structures requires integration times in a range of 45 to 125 milliseconds or less. This is not usually an overly strict restriction. For this "envelope" limit on analysis, assume the inter-mode frequency spacing is small. Therefore, the summation over modulations of different frequencies assumed in the term $E_{in} = e^{j2\pi f_n t}$ are nearly the same. Thus the time part of the phase was properly removed from prior equations of return from the target in the equations for $\Phi_e$ and $E$ in this report.

$$\left(f_j - f_i\right)t_{integrating} << 1 \quad \Rightarrow \quad e^{j2\pi f_i t} \approx e^{j2\pi f_j t} \tag{15}$$

As a comparison, start with the general rectangular plate solution of (10) for $M,N \geq 1$, but *only* activate the first mode along $x$, $m_{max} = M = 1$, in the integration of the return from the vibrating metal strip. Assign a cutoff mode defined by $\Delta z/2 = w_{max} = \Delta\varphi_n \sin \pi y/(b/n)$. Unless the modal distribution is uniform up to a cutoff mode, the sum would require modal participation factors $\Delta\varphi_{m,n}$ for each $w_{m,n}(x, y) = (\Delta z)_{m,n}(x, y) \rightarrow w_{1,n}(x, y)$. CSM concepts with (14) allow definition of a total phase change $\Delta\varphi$ which has units of length, to be divided by $\lambda$. An alternate MPF could have $\Delta\varphi' = (\Delta z)_{max}(x, y)/2\lambda N$, but to keep the optical wavelength exposed in following relations:

$$\phi_o = \frac{1}{MN} \quad (M > 0) \quad \xrightarrow{UMP} \quad \Delta\phi_{M=0} = \frac{(\Delta z)_{max}}{2N} \quad \Delta\phi_{M=1} = \frac{(\Delta z)_{max}}{2MN} \tag{16}$$

The overall displacement field $w(x, y) = \Sigma^{M,N} w_{m,n}(x, y)$ is the sum of all effective modal displacement vectors $\Delta\varphi_n \sin \pi y/(b/n)$. Here we dropped $m$ since $m_{max} = M = 0$ for a bar. $w$ represents actual displacement. It comprises a linear superposition of all the modes $N$ of a flat bar (where $M_{bar} = 0$). The exposed area $A = ab$. In (10), the vector $S_{avg}$ is a time averaged Poynting vector $\langle S \rangle = I$ and $c$ is the speed of light. Application of UMP, CSM, and especially SSS *allows the integrals and sums to be interchanged, which produces the Bessel function solution*:

$$\left|\vec{E}_{mod}\right|_{M=0} \cong \frac{2\sqrt{S_{avg}}}{c\varepsilon_o} = \frac{E_{in}}{ab}\int_0^b\int_0^a \exp\left[j\frac{4\pi}{\lambda}\left(\sum_{n=1}^N \Delta\phi_n \sin\frac{\pi y}{b/n}\right)\right]dxdy \xrightarrow{SSS} \frac{E_{in}}{b\pi^2}\int_0^b \sum_{n=1}^N \frac{n}{n} J_0\left(\frac{4\pi}{\lambda}\Delta\phi\right)bdy \tag{17}$$

Similarly for a single mode across the width, along x (where the cutoff x-mode is $M = 1$):

$$\left|\vec{E}_{mod}\right|_{M=1} \cong \frac{2}{c\varepsilon_o}\sqrt{\left|\vec{S}_{avg}\right|} = \frac{E_{in}}{ab}\iint \exp\left[j\frac{4\pi}{\lambda}\left(\sin\frac{\pi x}{a}\sum_{n=1}^N \Delta\phi_n \sin\frac{\pi y}{b/n}\right)\right]dxdy \tag{18}$$

The time modulation was removed by application of (15) assuming CSM in order to assume choice of mode shapes at the point in time when they are at maximum deflection, as is often customary for modal analysis.

### 3.0 Slowly Swept Sine (SSS) Spectral Elimination

The slowly swept sine (SSS) response is the scaled sum of a complete set of single tone responses where the nearest resonance to the driving frequency tone dominates. The response of other resonances is usually negligible. Although the integration over the optical spot size and mode summation (construction of structural deflection from active modes) will appear to swap, this is *not merely an interchange* of the integral and the sum of the modes in the exponent. Such an interchange would ordinarily be incorrect.[27] However, there is a physical linearity about resonance domination at the driving frequency that describes the physics of monotone FRF reconstruction; the integrand is a *sum* of many single mode responses. The form of the $M = 0$ return formula for the flat bar begs hopelessly for multi-modal simplification, which is not theoretically available, but we can force an SSS assumption. The area of integration $a \times dy_k$ over the segment $n_k$ changes with the chosen mode $k$. The scale for the field also changes using $\Delta\varphi_n = (\Delta z)_{max}(x, y)/2N$ in a uniform MPF assumption (UMP) where the "bandwidth" is N modes (see discussion before (5). So the field coefficient $E_{in}$ remains the sole coefficient. Ringing the system *one mode at a time*, the slowly sine swept (SSS) response is the sum of a set of single mode responses allowing the sum in the exponent to factor out as in (11) on page 22 provided UMP is appropriate. In this case the return is a *constant* for any mode, providing uniformity, SU. This estimate below is more carefully derived in the two following equations.

$$\left|\vec{E}_{mod}\right|_{M=0} \xrightarrow{SSS} \left|\vec{E}_{SSS}\right|_{M=0} \cong \frac{E_{in}}{b\pi^2} \int_0^b \left[ \sum_{n=1}^{N} \frac{n}{n} J_0\left(\frac{4\pi}{\lambda}\Delta\phi\right) b \right] dy = \frac{E_{in} N}{\pi^2} J_0\left(\frac{4\pi}{\lambda}\Delta\phi\right) = \text{const.} \quad (19)$$

Similarly, assuming a uniform MPF and normalized modes all set to $\Delta\varphi_n = \Delta\varphi = (\Delta z)_{max}(x, y)/2N$, application of SSS assumptions for (10) simplifies the relation. Because of the multi-modal nature of the underlying phenomenology, especially where modes are coupled, these SSS results do not provide a purely true modulated field at the detector. In simulation the Cross Spectral Covariance cross-terms can be substantial for both structural and propagated optical return.[3] Yet, as in practice, use of SSS can provide useful estimates and insight.

The integration scale change $\int_0^\pi f(\theta)d\theta = \int_0^\xi f(\varsigma)\xi d\varsigma/\pi$ introduced as $a/\pi$ (where $M = 1$) and $b/\pi n$ ($M = 0$) in the start of the SSS solution in the RHS of (17) and applied to (19), cancels the illumination area, $A = ab$. These factors turn the formulas into integrations over single half-cycles in $x$ and $y$ for $m, n > 0$. The Bessel function inside the summation in (19) seems desperate to jump out of the integration. However, if we look at what the terms represent, Guido Fubini's theorem[27] would appear to impede the simple interchange of this sum and integral.

Equation (20) confirms that the slowly swept sine (SSS) optical return from a bar is a constant over all modes. Here an expansion of the phase clarifies this mathematical observation. For a larger uniform modal participation "bandwidth" $N$, the average modal participation factor $\Delta\varphi_n = (\Delta z)_{max}(x, y)/2N$ decreases. This MPF turns out to represent an average range of deflection at a point due to the linear sum of all modes.

$$\left|\bar{E}_{\text{SSS}}\right|_{M=0} \cong \frac{2}{c\varepsilon_o}\sqrt{\left|\vec{S}_{avg}\right|} \cong \frac{E_{\text{in}}}{ab}\sum_{n=1}^{N}\frac{ab}{n}\frac{n}{\pi^2}J_0\left(\frac{4\pi}{\lambda}\frac{(\Delta z)_{\max}}{2N}\right) = \frac{E_{\text{in}}N}{\pi^2}J_0\left(\frac{2\pi(\Delta z)_{\max}}{\lambda N}\right) \quad (20)$$

It will be useful to repeat this calculation in still more detail. Bring the integral inside the sum and let the domain of the integral cover only one half-cycle $b/n$, multiplied by $n$ half-cycles along the $y$ direction: $\int_0^b \sum_{n=1}^N (\cdots) b\, dy = \sum_{n=1}^N \int_0^{b/n}(\cdots) b\, dy$ With this modification the interchange of sum and integral is no longer forbidden.[27]

Further simplification for the M = 1 case will require expansion of the Bessel function:

$$J_0(\xi) = 1 - \frac{\xi^2}{2^2(1!)^2} + \frac{\xi^4}{2^4(2!)^2} - \frac{\xi^6}{2^6(3!)^2} + \cdots \quad (21)$$

The plate response uses the expansion in (21) to modify (20), which is the expression for electric field modulated by a bar ($M = 0$), for the generic $M = 1$ case in order to represent return from plate vibration and other non-bar structures. This modification factors out one $(\Delta\varphi)^2$ providing a divisor of $N^2$ in (22) using the MPF described above, $\Delta\varphi = (\Delta z)_{\max}(x, y)/2N$.

$$\left|\bar{E}_{\text{SSS}}\right|_{M=1} \approx \frac{E_{\text{in}}}{b}\pi\sum_{n=1}^{n_{\max}}\frac{n}{\pi}\int_0^b J_0\left(\frac{4\pi}{\lambda}\Delta\phi_n \sin\frac{\pi y}{b/n}\right)dy$$
$$\approx E_{\text{in}}\left[\left(\sum_{n=1}^{1}1\right) - \left(\frac{4^2\pi^2}{\lambda^2}\frac{(\Delta\phi)^2}{4}\frac{1}{2} + \frac{4^4\pi^4}{\lambda^4}\frac{(\Delta\phi)^4}{64}\int_0^b \sin^4\frac{\pi y}{b/n}dy - \cdots\right)\left(\sum_{n=1}^{N}1\right)\right] \quad (22)$$

An identity (37) for integrals of even power of sine simplifies integrals distributed into the expansion of $J_0(\xi)$. Consider $(\Delta\varphi)^4$ and higher order terms negligible. The structural wavelength $2b/n$ decreases with increasing mode number (and vibration frequency). However, the constant term in front, the leading '1' in the Bessel function expansion in (21), comprises only one mode and thus its maximum mode number $n_{\max}$ is unity, not $N$. See the first term of (22) where the sum is over a single term compared to the counting sum to $N$ at right which is a coefficient for the rest of the terms. This leading term will come to represent the reflected unmodulated return. $(\Delta\varphi)^2 = (\Delta z)_{\max}^2/4N^2$ factors out of all of the modulation terms in the series in (22) so that each $\eta$'th term is proportional to $((\Delta\varphi)^2)^\eta$ starting with a constant term ($\eta = 0$).

Because of the two dimensional nature of the plate vibrations, even though the $x$ dimension is limited to the $M = 1$ cutoff mode, the argument is no longer strictly a constant, but rather a harmonic function of $y$: $\sin \pi y/(b/n)$. Thus the expansion (21) was necessary to obtain a "first order term" inside the last part of (22). This "un-modulated return" value for part of the sensed field is a partial validation of the suitability of form of the equation.

## Non-Imaging Detected Signal, Numerical Integration

The SSS general UMP plate return for $M = 1$ in (22) is not as simple as the flat bar return ($M = 0$). However, the numerical work to estimate results for a particular structure does not require simplifications such as the interchange of the sum and the integral[27] so that we can use the starting point, (9) from page 21 copied here:

$$\vec{E}\Big|_{M,N \neq 0} = \int_0^a \int_0^b \frac{\vec{k} \cdot \vec{r}}{kr} \frac{E_{in}}{A_{ab}} \exp\left[-j \frac{4\pi}{\lambda_{opt}} \left(R + \sum_{m,n=1}^{M,N} (\Delta z)_{m,n} \sin\frac{\pi x}{a/m} \sin\frac{\pi y}{b/n}\right) + j\phi_r\right] dxdy \quad (23)$$

Numerically summing up the modal modulation[3] with $M = 0$ in the initial modulation relation (23) we obtain a constant total radiant flux $\Phi_e(t)$ for a bar. For $M = 1$ a wide bar or plate is probably held or simply supported on all edges and in this case has stiffness against bending about the *y*-axis perhaps merely due to a shorter width which excludes modes with higher bending – the stiffness excludes smaller spatial wavelengths along *x* greater than half the width along *x*. This $M = 1$ return field from such a plate results in an argument in the Bessel function for the expression for $\Phi_e(t)$ that varies with location in y. In the plot of the results of these equations (Figure 9), this variation is apparent in the values representing modulation due to bars. Plane Wave SSS $\Phi_e$(mode) return plotted for each mode *N* due to spatially integrated (non-imaging) 2-D plane wave solutions (filled circles ●) varies with mode number. In contrast the SSS signal received from strips or bars does not vary (values are in the centers of the squares □, which align to the same point); optical SSS return from 1-D vibrations is constant over all modes at this M = 0 extreme. The ever-constant summand in (20) representing optical return from one Chladni zone for any mode *n* of a strip (bar) is the set of superposed squares [□] in Figure 9. They represent several solutions with the same exact result (mode *n*=0, $\log_e \langle \Phi_e \rangle \approx -14.4$). So for vibrating reflecting strips (bars), within the 'Assumptions of the Plane Wave Solutions' (page 19), the SSS modulated return remains constant over all modal variations due to the free-free BC's on both sides. In contrast, the variation of two-dimensional plane wave solutions even for unnaturally high symmetry given by (23) for return from a rectangular panel (displayed in the plot as filled-in circles ●) indicates that far–field CW application will have enough SSS variation to classify different vibration mode shapes.

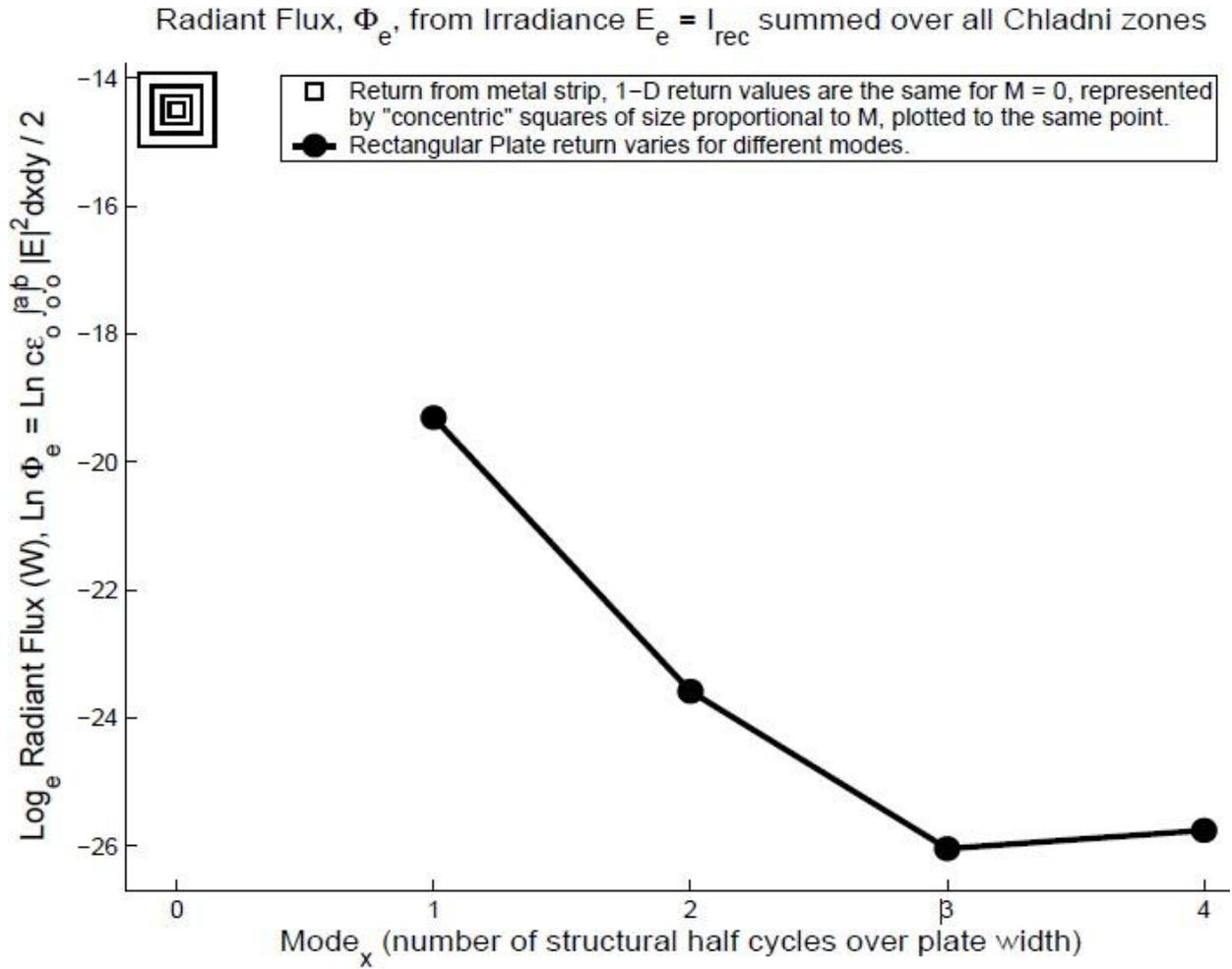

Figure 9.  Non-imaging SSS return from large spot size probe on bars □ versus plates ●

Analytical SSS approximation

The result in (24) below more easily compares with the MM result in the next section (38), as does (25) with (40).  Since targets usually have $N > 2$, this approximation for $M = 0$ is within *engineering error*[††] for deflections up to the order of a wavelength of vibration amplitude $(\Delta z)_{max} < 2\lambda$:

$$\left|\bar{E}_{SSS}\right|_{M=0} \approx E_{in}\frac{N}{\pi^2}J_0\left(\frac{2\pi(\Delta z)_{max}}{\lambda N}\right) = E_{in}\frac{N}{\pi^2}\left[1 - \frac{1}{4}\frac{4\pi^2(\Delta z)_{max}^2}{\lambda^2 N^2} + \cdots\right] \approx E_{in}\left[1 - \left(\frac{(\Delta z)_{max}}{\lambda}\right)^2\frac{\pi^2}{N}\right]$$
(24)

---

[††] Engineering Error is generally up to 5-10% but in this case the remainder is smaller.

Using similar methods and with the expansion of the Bessel function in (21), the SSS (20) evolves into the $M = 1$ slowly sine swept return:

$$\left|\vec{E}_{SSS}\right|_{M=1} \approx E_{in}\left[1 - \left(\frac{4^2\pi^2}{4\lambda^2}\frac{(\Delta z)^2_{max}}{4\pi N^2}\frac{\pi}{2} + O[(\Delta z)^4] - \cdots\right)N\right] \approx E_{in}\left[1 - \left(\frac{(\Delta z)_{max}}{\lambda}\right)^2\frac{\pi^2}{2N}\right] \quad (25)$$

The larger coefficient of the first order modulation term on the RHS of (25) compared to (24) is not as different as it appears, $\pi^2/2 \approx 4.935$ compared to $\pi^2$. The structural dynamics of the system where the length is greater than the width ($b > a$) establishes that the cutoff mode along $y$ is far higher when another $x$-mode becomes active, by over an order of magnitude. An over-simplified "wide" beam first order estimate could be useful to see this result. Comparison of the $y$ cutoff $N$ for (25) which is $M = 1$, to $N$ in (24), which is $M = 0$, will show that $N_{M=0} \gg N_{M=1}$. Consideration of the behavior of vibration bending modes in a flat plat approximated by the beam equation will show that the first order corrections for the two cases are not as different as they appear.

It is possible to estimate of the extent to which the fundamental cross-mode of vibration across the width of a bar (slender plate) absorbs energy only when a large number of modes along the length are already active. From the typical bending stiffnesses via the deflection equation for a simple cantilever beam[31] it is apparent that the frequency causing modulation along the $y$-direction (bending about $x$) is proportional the frequency causing modulation along the $x$-direction (bending about $y$) by a factor (length/width)$^2$ = $(b/a)^2$ which is the square of the aspect ratio or inverse thereof. There is no practical formula for plate vibration because the eigenvalues are a function of the aspect ratio itself, each a different eigen-solution and for different boundary conditions (fixity). As a rough estimate the prior result that $f_{yy} = f_{xx}(b/a)^2$, combined with the numerically calculated frequency coefficients for the first three modes of the same simple cantilever beam,[32] provides an expression for the aspect ratio at which higher level beam modes of $f_{xx}$, herein $f_{n=1}, f_{n=2}, f_{n=3}$, will match the fundamental cross-mode $f_{yy}$ for $M = 1$ which undulates along the $x$ direction ($w[x] = A'\cos[k_x x]$). The rotation *about* the $y$-axis drives the mechanical engineering terminology in double subscripts like '$yy$.' Define a coefficient of the modes $q$ in terms of Thomson's parameter $\beta_n$ and the beam length $l$: $\zeta \equiv (\beta_n l)^2$. Selecting the values for a cantilever beam:[32] $\zeta = [3.54\ 22.0\ 61.7]$. The square root of these ratios provides the aspect ratio at which the cross-beam fundamental mode $f_{yy}$ (which undulates along $y$) will match a higher mode: aspect$(f_{yy} = f_{n=2})^2 = (b/a)_{M=1}^2 = \zeta_2/\zeta_1$ which is $(b/a)_{M=1,n=2} = 2.50$ (strangely enough, *exactly*). Similarly $(b/a)_{M=1,n=3} \approx 4.19$. Recall that a beam or bar is a slender object where this (inverse) aspect ratio is approximately infinity, or using the $x$ direction width divided by the length, $(a/b)_{bar} \ll 1$. Therefore, as is usually seen in practice, typical beam-like structures such as a foot long plastic ruler will increase the $y$ direction mode number $n$ up to a cutoff of approximately $N = 12$ before $m = 1$ becomes active at cutoff $M = 1$. Therefore, in comparison of $N$ for (25), which is $M = 1$, to the $y$ cutoff $N$ in (24), which is $M = 0$, it is reasonable to expect $N_{M=0} \gg N_{M=1}$. So the cutoff $N$ in (25) must be much larger than the cutoff $N$ in (24). Thus the first order corrections for the two cases, constant deflection along $x$ and one half-cycle over the width, are not as different as they appear.

There is a situation in electro-optics analogous to SSS. Derivation for monochromatic phase modulation in a crystal provides a similar result[26] because the argument of the Bessel function is small. The referenced text describes optical modulation using birefringent material properties represented by the electro-optic (EO) coefficients $r_{i,j}$ which represent the effect of the index of refraction ellipsoid ($n_o$, $n_e$) of the device of thickness $d$ on electric field components $E_m$. In that reference, Yariv and Yeh's phase-modulation index $\delta = \pi n^3{}_o r_{63} E_m d/\lambda_s$ is similar to the effect of the deflection at one small spot size point on the target in wavelengths, except that they investigate the temporal variation in phase. Therefore, for a laser frequency of $\omega = c/2\pi\lambda$ we define in (26) a phase-modulation index for the laser vibrometry problem. There is a variation in phase over space as deflection varies for each mode vibrating at a frequency and $\omega_{m,n} = 2\pi f_{m,n}$ as in Yariv and Yeh:[26] $E_{out} = A_o \cos[\omega t + \delta \sin(\omega_{m,n})]$. The interferometric modulation the index $\delta$ becomes:

$$\delta_{mod} = \frac{2\pi w_{m,n}(x,y)}{\lambda} = \frac{\pi (\Delta z)_{m,n}(x,y)}{\lambda} \xrightarrow{UMP} \delta_{mod} < \begin{cases} \left(\frac{(\Delta z)_{max}}{\lambda}\right)\frac{\pi}{N} & M = 0 \\ \left(\frac{(\Delta z)_{max}}{\lambda}\right)\frac{\pi}{2N} & M = 1 \end{cases} \quad (26)$$

## 4.0 Multi-Modal (Small W) Spectral Reduction

In (9) (copied on page 28 as equation 23) the integration over the bar length using the large spot size, non-imaging, UMP and CSM assumptions for the slowly swept sine (SSS) case created Bessel function solutions. Stopping short of this integration, an expansion of the field integrand using a truncated series approximation for the real part of Euler's relation, $\mathcal{R}e\,[e^{j\theta}] = \cos\theta = 1 - \theta^2/2! + \theta^4/4! - \ldots$ generates another useful estimate, the small deflection *multi-modal* (MM) estimate. This "cosine" term is from the real part of the analytic form of modulation of the received signal (shown in a second copy of (9) or (23) from pages 21 and 31)):

$$\vec{E}\Big|_{M,N\neq 0} = \int_0^a\int_0^b \frac{\vec{k}\bullet\vec{r}}{kr}\frac{E_{in}}{A_{ab}}\exp\left[-j\frac{4\pi}{\lambda_{opt}}\left(R + \sum_{m,n=1}^{M,N}(\Delta z)_{m,n}\sin\frac{\pi x}{a/m}\sin\frac{\pi y}{b/n}\right) + j\phi_r\right]dxdy \quad (27)$$

The following MM result using this expansion to $\theta^2$ order provides insight into the participation of each mode in the optical return. This approximation assumes deflections are very small ($2w(z) = \Delta z < 8\,\mu m$) in order to omit higher order terms. Unlike estimates in the previous case, a MM result is not due to monotone excitation but fully coupled multi-modal response. Again, the 1-D vibration mode assumption can simplify the return field calculations. So for $M = 0$ where the $x$ direction is already integrated out ($\times a$) the optical return comes from a specular strip or bar that has zero cross-modes across the width. Equation (28) follows directly from (27) by use of the second order of the small deflection based function $\theta$ just described above:

$$\Big|\vec{E}_{MM}\Big|_{M=0} = \frac{E_{in}}{b}\int_0^b\left(1 - \frac{1}{2}\frac{(4\pi)^2}{\lambda_{opt}^2}\left(\sum_{n=1}^N(\Delta\phi)_n\sin\frac{\pi y}{b/n}\right)^2 + O\left(\left[\frac{(\Delta z)_{max}^2}{\lambda_{opt}^2}\right]^2\right) + \cdots\right)dy \quad (28)$$

The first step is to expand the Euler exponential into a power series of the real part of the same exponent which comprises the $\theta$ term from $\mathcal{R}e\,[e^{j\theta}]$. As accomplished for the previous method (SSS) this MM estimate chooses an arbitrary phase to cancel the $\varphi_r$ in (27) and $\exp[j\pi R/\lambda]$ is dropped from calculations. The complete phase term $\theta$ described above contains the sum of sinusoidal modes and the deflection and diffraction coefficients (including $\Delta z$ and $1/\lambda$). The third-order terms of the series in (28) are equivalent to $O((\Delta z_{max})^4/\lambda^4)$ for the *uniform* modal participation (UMP) model. They are much smaller than the second-order terms bound below by $O(\Delta\varphi_n^2/\lambda^2) = O((\Delta z_{max})^2/\lambda^2)$. From (28) introduce $(\Delta\varphi)_n$ and use $(\Delta\varphi)_n \equiv (w_n)_{max}/N$ in (29) with the UMP value MPF $\equiv (\Delta\varphi)_n = \Delta\varphi$ (a constant). Again, the calculation uses a cutoff mode $N$ which defined a form of bandwidth. Thus the sum in (28) simplifies to represent the maximum *average* composite (system) deflection at $y$:

$$\sum_{n=1}^{N}(\Delta\phi)_n \sin\frac{\pi y}{b/n} = \sum_{n=1}^{N}\frac{w_n(y)}{N}\sin\frac{\pi y}{b/n} \xrightarrow{\text{UMP}} \sum_{n=1}^{N}\frac{w_{avg}(y)}{N}\sin\frac{\pi y}{b/n} = \frac{\Delta\phi}{N}\sum_{n=1}^{N}\sin\frac{\pi y}{b/n}$$
(29)

The right hand side (RHS) of (29) shows that since $N$ cannot be too small,[‡‡] an averaged maximum of system deflection, the sum of harmonics, is small all along the length, as are the actual stable deformed shapes. The deflection in (30) defines the MPF, $\Delta\varphi/N$, a constant, where the term at right is not near the end effects seen in Figure 10(a) for $y < 2b/n$ (as discussed later).

$$\frac{(\Delta z)_{max}(y)}{2N} = \frac{w_{max}(y)}{N} > \sum_{n=1}^{N}\frac{w_{avg}(y)}{N}\sin\frac{\pi y}{b/n} = \frac{\Delta\phi}{N}\sum_{n=1}^{N}\sin\frac{\pi y}{b/n} \qquad y > \frac{2b}{n} \quad (30)$$

This UMP condition *appears* to be a Fourier series of a constant $\Delta\varphi/N$ with kernel $\sin \pi y/(b/n) = \sin[n(\pi y/b)]$. It might transform the "spatial spectrum" $\Delta\varphi/N$ as $\mathcal{F}_{\sin}[\Delta\varphi]$, except the integral of the sum *implies* an inverse transform[33] to produce the "signal" which is the phase $\theta'[k]$.

$$\theta'[k] \equiv \Delta\phi\, j\frac{1}{N}\sum_{n=1}^{N}\sin\left(\frac{2\pi k n}{N}\left[\frac{yN}{2bk}\right]\right) = \pm\Delta\phi\left[\frac{yN}{2bk}\right]\delta\left(\left[\frac{yN}{2bk}\right][k - k_{yo}]\right) \qquad (31)$$

Ordinarily the Fourier sine transform of a constant is an odd function, two opposing delta functions (spikes) at the pertinent spatial frequencies, $\pm k_{yo}$. In spite of its form in (31), it would be proper to resist the temptation[§§] to see what phase term modulates the reflected probe beam, the imaginary inverse discrete Fourier sine transform producing a real odd signal $\theta'(\pm k_{yo})$. This sub-analysis of the phase would tend to validate the UMP assumption but the real dynamics are a bit more subtle. After a cacophony of algebraic simplifications, without the previous SSS assumptions, the numerical analysis below summarized in Figure 10 provides a better validation.

---

[‡‡] Not only does common practice prove $(w_n)_{max}/N$ can be small, the smallest the denominator $N$ can be is unity. Therefore, small deflection assumptions hold; $E_{MM}$ of (28) cannot be singular due to small $N$, even if the inverse appears to grow as $N$ decreases, because in this application of an inverse, $N$ cannot be less than unity (1).

[§§] Equation (31) is the inverse discrete Fourier transform (IDFT) for a real and odd sequence, because the cosine terms for such a sequence have coefficients that are all zeros. It is a finite sum, $n = 1$ to $N$, of the sort used to transform finite sequences. The ordinary form assumes the argument of the sine is an integer scaled only by $2\pi/N$ which is the reason the scale property of the Fourier transform appears in the result on the right of (31).

It may be useful at this point to recall the framework. $\Delta z(y)$ is half the spatial *amplitude* $w(y)$. At each location $y$ along the bar, the oscillations in time cycle through the undeformed shape and then out to a maximum at $\Delta z(y)/2$. This term is both a spatial and temporal amplitude.

The next step is to solve (28) explicitly while ignoring terms of order $O([\Delta z/\lambda]^4)$ or smaller values, dropping the 'higher order terms.' The correction terms in (28) now contain *a square of a sum*. This is an exploitable mathematical form. Solution of this equation 'to second-order' makes use of the series solution for each term in the expansion of the *square of a sum* of ordered sine functions $\psi_k = \sin(y k n_k)$ such that the square of the sum can be separated into a sum of squares and a cross-term sum (which sums to a convenient value):

$$\left(\sum_{p=1}^{N}\psi_p\right)^2 = \sum_{p=1}^{N}\psi_p^2 + 2\sum_{p\neq q}^{N}\psi_p\psi_q \tag{32}$$

(Expand $(x+y+z)^2$ to see this.) Specification of the functions $\psi_k = \sin(y k n_k)$ into (32) would require different $n \to n_1, n_2$ and $y \to y_1, y_2$ in order to distinguish $\psi_p$ from $\psi_q$. To be strict, the $y$ location is the same variable so we can drop location subscripts $y_1, y_2 \to y$. But the modes must differ. Next, allow this derivation to avoid this cross-term subscripted algebra with another variable re-definition, among others. Integrating the second term in (28) along the beam length $y$,[34] for modes that have BC's held or simply supported at the edges,[31] and specifying different $y$-direction mode numbers $n_1, n_2 \to p, q$, a simplification appears. In the temporary $p, q$ shorthand in (33), and then in (34) the cross terms $\psi_p\psi_q$ become an expression of cosines.

$$\psi_p\psi_q = \sin\frac{\pi y}{b/p}\sin\frac{\pi y}{b/q} = \frac{1}{2}\cos([q-p]\pi y/b) - \frac{1}{2}\cos([q+p]\pi y/b) \tag{33}$$

When the $y$ variable is shifted to the plate center, $y \to y' - b/2$, the cosine expressions in (33) become odd functions. They integrate to zero directly here. The formulas in the standard integration tables[34] and (34) verify the disappearance of the cross-terms by different methods:

$$\int_0^b\sum_{p\neq q}^{N}\sin\frac{\pi y}{b/p}\sin\frac{\pi y}{b/q}dy = \left[\frac{\sin([q-p]\pi y/b)}{2[q-p]} - \frac{\sin([q+p]\pi y/b)}{2[q+p]}\right]_0^b = 0 \tag{34}$$

Since $y_{\max} = b$ is a held edge BC,[31] *all of these cross terms cancel* to produce the property that the integral of the square of these particular sums equals the integral of the sum of the squares. Therefore substitution of (32), using the resulting zero in its second term, and distributing the integral onto the second order (modulation) term in (28), the integral of the *square of the sum* becomes an integral of the *sum of the squares*. This last term has an easy solution.

$$\int_0^b\left(\sum_{p=1}^{N}\psi_p\right)^2 dy = \int_0^b\left(\sum_{p\neq q}^{N}\sin\frac{\pi y}{b/p}\right)^2 dy = \int_0^b\sum_{p\neq q}^{N}\sin^2\frac{\pi y}{b/p} + 0 \tag{35}$$

With this result from (34), (35) provides the main simplification for the expression for the multi-modal return, (28). The uniform participation model (UMP) allows the sum in expression for modulated electric field to factor and simplify, resulting in a single "correction" term for the second-order calculation of (28). Note that the derivation followed the same path as (17) except for the MM assumption rather than SSS,

$$E_{M=0} \Rightarrow \underbrace{\int_0^b \frac{E_{in}}{b} dy - \frac{E_{in}}{b}\left(\frac{4^2\pi^2}{\lambda^2}[\Delta\phi]^2\right)\int_0^b \left(\sum_{n=1}^N \sin\frac{\pi y}{b/n}\right)^2 dy}_{MM} \Rightarrow \underbrace{E_{in} - \frac{E_{in}}{b}\left(\frac{4\pi\Delta\phi}{\lambda}\right)^2 \sum_{n=1}^N n\int_0^b \sin^2 \frac{\pi y}{b/n}\frac{\pi dy}{b}}_{Eq.35}$$

(36)

Equation (23) results in an unmolested term ($E_{in}$) and a term that represents the variation of the return due to the structural vibration. Since it is appropriate to bring the integral inside the sum in this variation term,[27] each summand integrates to the same value:[34]

$$\int_0^\pi \sin^2(n\theta)d\theta = \frac{\pi}{2} \qquad \rightarrow \qquad \frac{b}{\pi}\int_0^{b/n} \frac{\pi}{b}\sin^2\frac{\pi y}{b/n}dy = \frac{b}{2}$$

(37)

Equation (37) uses a generic integral identity to expand from $n = 1$ half-cycles to different mode indices $n > 1$ where $\theta \equiv \pi\xi/b$ and thus $\theta \equiv \pi\xi/ bd\theta = n\pi d\xi$. To calculate modulation over one Chladni zone, the spatial integration over $\theta$ from 0 to $\pi$ via (37) becomes $ab/2$. The RHS of (36) gives the result for one Chladni zone and then sums the number of identical zones in each mode, which is the mode number in this 1-D vibration state (for a flat bar). This solution expression requires further simplification provided by the UMP assumption, a uniform MPF distribution $\Delta\varphi = (\Delta z)_{max}(x, y)/2N$ for all structural modes n. In the first term $b$ will cancel because the field density over area of width $a$ is ($a \times$ length) where 'length' is the half-cycle expression for sine squared integration on the RHS of (37). Therefore, the effective area is $ab/2$. $b$ from this area cancels that in the denominator preceding the integral, RHS of (36), leaving a factor of ½ seen in (38). Similar to a "cutoff frequency," the highest mode number $N$ provides an estimate of vibration strain energy bandwidth in low-frequency modes.

$$\left|\vec{E}_{MM}\right|_{M=0} \approx E_{in} - \frac{1}{2}\frac{4^2\pi^2}{ab}E_{in}\frac{(\Delta\phi)^2}{\lambda^2}a\left(N\frac{b}{2}\right) = E_{in}\left[1 - 4\pi^2\left(\frac{(\Delta z)_{max}}{2N\lambda}\right)^2 N\right] \approx E_{in}\left[1 - \left(\frac{(\Delta z)_{max}}{\lambda}\right)^2 \frac{\pi^2}{N}\right]$$

(38)

Presuming UMP and the first MM approximation, very small deflection $(\Delta z)_{max} \ll \lambda\sqrt{8/\pi^3}$ $\approx 0.508\,\lambda$, (38) shows to second order that the deviation from a response from a flat plate with no vibrations (the first term) is inversely proportional to the cutoff mode number $N$. The wider the response bandwidth, the less improvement adding another mode provides for the estimate. Physically, this insensitivity to cutoff number $N$ (insensitivity to response with high frequency content) is an effect of this being a one-dimensional displacement field.

The 2-D form restricted to on $x$-direction half cycle, $M = 1$, starts with an intermediate form between (27) and (28) where the $x$ direction is not yet integrated out. Thus assignment of $M = 1$ will modify the correction term, using (37) to add a factor of $a/2$. The modulation terms $E_{\text{mod}}$ will be similar to the second term on the RHS of (38). Assign a geometric field density $E/\Delta A$ over area in (39). The displacement term in the UMP expression appears as a ratio of dots. This provides a field density of $E_{\text{in}}/4$ if only a single $y$-mode is active, in addition to the lone $x$-mode. The actual structural displacement field is a linear combination of this one mode along $x$ and all the $N$ modes in $y$.

$$\left.\frac{|\vec{E}_{\text{mod}}|}{\Delta A}\right|_{\substack{M=1 \\ MM}} \Rightarrow \frac{1}{2}\frac{E_{\text{in}}}{\Delta A}\left(\frac{4\pi\Delta\phi}{\lambda}\right)^2 \left[\frac{a}{2}\right]\frac{b}{\pi}\int_0^{b/n}\frac{\pi}{b}\sin^2\frac{\pi y}{b/n}dy = \frac{E_{\text{in}}}{2ab}\left(\bullet\right)^2\left[\frac{a}{2}\frac{b}{\pi}\frac{\pi}{2}\right] = \frac{E_{\text{in}}}{8}\left(\bullet\right)^2 \tag{39}$$

Since only one $x$-mode exists, its integral produces an extra ½ coefficient in the result $a \rightarrow a/2$. Therefore, the $x$-direction $M = 1$ modulation results in an overall factor of ½ which carries to the last term where the sum and integral are switched, and ultimately to the estimation formula in (40). The result is that with only one mode across the width ($M = 1$) the second return correction is inversely proportional to $N$ (but only by half as much as in the $M = 0$ case):

$$\left.|\vec{E}_{\text{MM}}|\right|_{M=1} \approx E_{\text{in}} - \frac{1}{2}\frac{4^2\pi^2}{ab}E_{\text{in}}\frac{\phi^2}{\lambda^2}\frac{a}{2}\left(N\frac{b}{2}\right) = E_{\text{in}}\left[1 - 2\pi^2\left(\frac{(\Delta z)_{\text{max}}}{2N\lambda}\right)^2 N\right] \approx E_{\text{in}}\left[1 - \left(\frac{(\Delta z)_{\text{max}}}{\lambda}\right)^2\frac{\pi^2}{2N}\right] \tag{40}$$

The $M = 0$ and $M = 1$ and cases compare to each other directly:

$$\left.|\vec{E}_{\text{MM}}|\right|_{M=1} \approx \frac{1}{2}\left(E_{\text{in}} + \left.|\vec{E}_{\text{SSS}}|\right|_{M=0}\right) \tag{41}$$

These two MM results match the SSS results; (24) and (25) provide the same first order deviations of sensed field compared to the $M = 0$ and $M = 1$ equations shown above even though the MM method used here is an entirely different calculation. Within these MM assumptions the x dimension modulation restricted to $M = 0$ (25) or $M = 1$ (26) provides response which *exhibits insensitivity to high frequencies*. The correction term for $M = 1$ is larger but it is negative, providing a smaller return for the same number of modes, as expected.

Extending to practical structures where $M > 1$ (and $N > 1$) solutions are more complicated and do not provide simple monotonic results (such as the correction term proportional to $1/N$ in (40)) since integrals of all orders and all combinations do not allow cancellation in the sum used in the transition from (28) to (36) to (40). **This allows for spectral classification, even for perfectly rectangular wide plates** partly because for $M > 1$, modulation term $\psi_p$ is more complicated, (36) no longer allows sums of squares to factor out via $\int_0^b \left(\sum_p^N \psi_p(ky)\right)^2 dy$.

The lack of SR and SU a necessary but not sufficient condition for acceptance criteria of CW return (CSC's,[3] spectra, and MAC's[15]) to be adequate classifiers for actual rectangular plates. These calculations show vibration classifiers built on large spot size return are incomplete classifiers for metal strips (bars) due to SR, yet adequate for structural ID of even perfectly rectangular plates where necessarily M > 1.

Figure 10 below contains a lot of information defined over the next several pages. The upper pane shows how several low frequency modes sum together into an edge concentration much higher than a Gibbs 7% horn. The non-imaging return uses these values of deflection for its phase modulation via the sum of all modes $n$ up to the cutoff $N$ where the total deflection appears in picometers on the ordinate. The deflection over the spatial extent (the abscissa) contributes to the return that the detector sees in order to obtain the radiant flux as a function of time, $\Phi_e(t)$ as a sum over $N$ modes. The "carrier frequency" time oscillation was omitted from the calculations back in the discussion of Figure 6 and the later CSM discussion using (15). However, time modulation is implicit in the concept of modes where we might register the wavelength seen in the upper pane instead of the structural vibration frequency where that mode freely vibrates and into which it receives driving vibration strain energy.

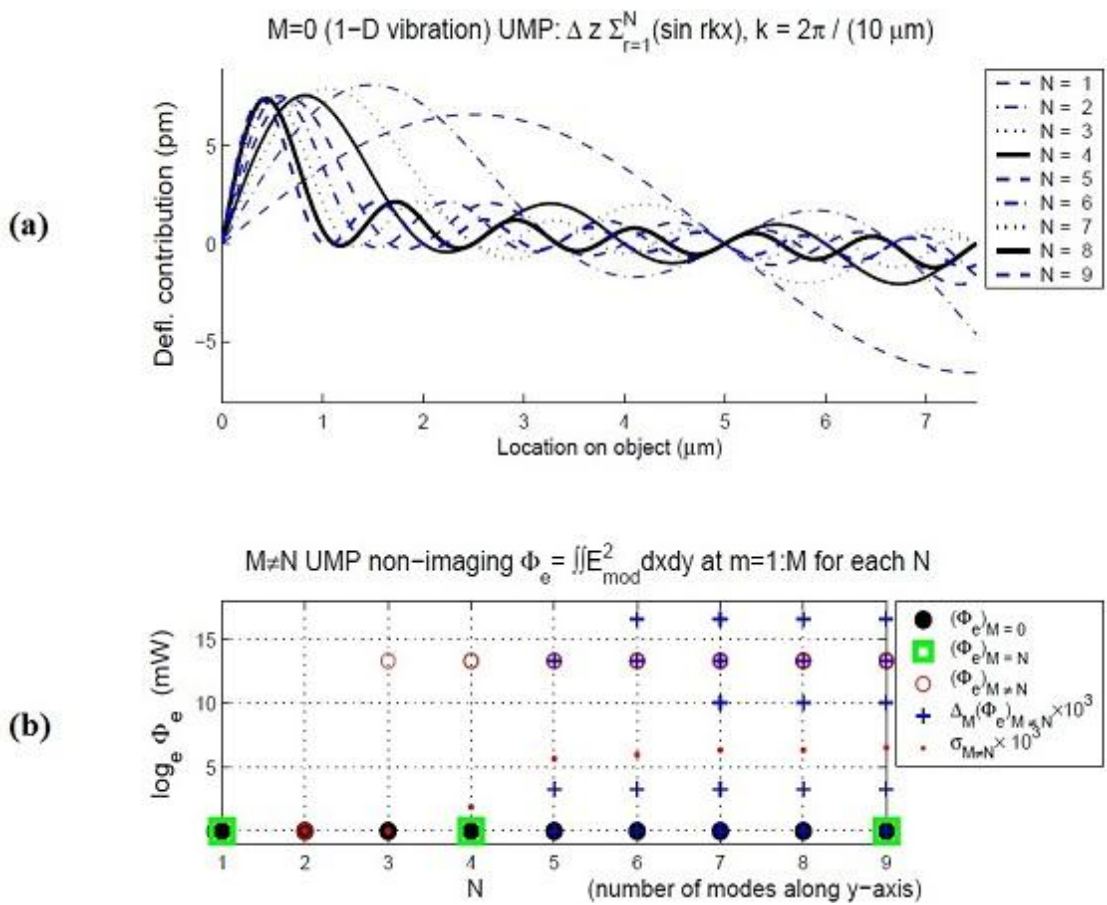

Figure 10. (a) Multi-mode small deflection stack-up, (b) Non-imaging irradiance by max mode $N$.

*Without using the approximations* of (36), the Matlab plot in Figure 10(b) shows that the 1-D non-imaging large spot size return is theoretically indistinguishable for any number of modes, to within machine error. This constant value for all the radiant flux from a bar (filled circles ●) shows that for $M = 0$ the modulation terms in (28) are negligible; $\Phi_e$ is not a function of $N$. By brute force, without using the small deflection assumption, the modulation terms for $M = 0$ sum to zero within machine error tolerance.

In comparison to the nature of the response for $M = 0$ results estimated in (40) (●), the $M = 1$ terms show a similar spectral uniformity effect not yet validated in measurements (the □ at zero phase for $N = 1$). The MM calculation for Figure 10(b) uses an $R = 0$, $\varphi_r = 2\pi n$ form of (27) described in the introduction. In reality the manufacture of even a flat square plate with $M = 1$ *would require considerably exotic fixtures to constrain all higher x-axis modes ($M > 1$)*. Thus, Figure 10(b) implies that for primitive flat structures that are *perfectly* super-symmetric structures (both have $M < 2$), the excitation of more than a few modes can make little difference for spectral ID; the return is insensitive to modes of higher frequency, they are SU. As discussed below, higher cutoff systems, or more *natural* structured systems ($M > 1$), are not.

The results for $M = N$ and $M \neq N$ (□ and ○) have more detail that deserves more discussion.

## 4.1 Details of the MM Un-approximated Plot

Figure 10(a) shows the deflection shape for very high mode counts becomes overly high near the edges, which is where the uniform modal participation model breaks down if deflection is not very small. The spatial extent of several modes that sum together produce the radiant flux shown in the lower pane, plotted versus the number of modes summed together. Each half cycle spans the length of the Chladni zone for that mode. In this case the first mode, $n = 1$ has a half-wavelength span of 5 microns. Higher frequency modes have progressively smaller zone lengths. This effect is similar to the frequency domain physics of mode-locked lasers where the etalons interact to form frequencies of strong laser feed stock. The field $E$ has an expression identical to that used to calculate values in Figure 10(a).[28] Laser engineers calculate the finesse of the edge spikes, the ratio of free spectral range $v_{fsr}$ to the transmission line width $v_c$ as a function of resonator reflection radii.[28] For structural mode stack-up the geometry of the boundary conditions (BC's) of the edges of vibrating panels, if such modification is possible, accomplishes the same mode control. For water waves spill edges for swimming pools accomplish some control of these modal superposition issues which are pronounced just away from the pool edge on the sides parallel to the lanes, as Figure 10(a) implies.

These laser vibrometry results in Figure 10 are useful for small deflection where the common modulation term $O(\Delta\varphi_n^2/\lambda^2)$ is the largest modulation term because $O(\Delta\varphi_n^4/\lambda^4) = (\pi\Delta z_{max}/\lambda)^4/N^2$ « 1 is small, so that deflections are small enough compared to the probe laser wavelength. This condition is equivalent to $\Delta z_{max} \ll \lambda\sqrt{N}/\pi$ which for a 10 micron laser using four structural vibration modes for classification restricts the utility of the second order formula in (28) to $\Delta z_{max}$ « 6 microns. Thus, a combination of uniform $\Delta\varphi$ and an estimated modal participation cutoff past the fundamental mode (the uniform spectral mode energy "distribution" assumption, UMP) provides a modest theoretical basis for a sufficient (not necessary) condition for the spectral elimination seen in the prior Doppler lab measurements and the qualitative explanation in that thesis.[2] Taken to the full illumination extreme (large spot size), if deflections remain small, additional modes provide little change in the overall return. This is an example of spectral uniformity (SU) for "high" frequency modes.

Allowing for structural variation, violation of the assumptions of UMP, imperfectly linear sides and mode shapes and material properties, and imperfectly flat undeformed shapes, the ability to distinguish 1-D mode shapes becomes progressively better. Modeling return from square 2-D symmetric modes $M = N$ in Figure 10(b) presents a borderline case of modal reduction for cross modulation that is similarly unlikely to be found in practice. Plotted as square symbols □ in Figure 10(b) this data is from the product of two sine functions representing the Chladni zone – vibration of $x$ and that of $y$, both inside the cosine term ($\mathcal{Re}\,[e^{j\theta}] = \cos\theta = 1 - \theta^2/2! + \theta^4/4! - \ldots$). When scaled up by 1,000 the values zoom up to a visible variation which can be seen by looking at the variance of radiant flux $\sigma$ shown as small red dots · in the lower pane of Figure 10. Thus the *SU is borderline and restricted to super-symmetric perfect structures*.

This report responds to the concept that SE may be an issue in laser vibrometry. The analysis finds that the confluence of modulation, spatially integrated across the profile, indicates an expected SU difficulty for laser vibrometry to classify return from vibrating strips of different lengths and perfectly square plates. For this super-symmetry the non-imaging return $\Phi_e$ (in watts) does not vary per mode as does $\Phi_e$ from ordinary structures. Detection is susceptible to illumination coverage of pairs of Chladni zones whose reflected exitance interferes with one another producing SR. With less perfect structures the collapse of the SR symmetries reduces cancellation. This reduction of destructive interference provides a variation in return shown in the two dimensional flat armor plate example[27] displayed in simulation (Figure 4 on page 8) which provides a spectral identification classification capability. Image streams with images arranged as in Figure 4 (right), time histories of simulations,[3] and the resulting cross-spectral covariance plots[3] show clear resonances at the structural modal frequencies. Those are the kind of low frequency resonances that should be straightforward to detect with spectral ID of return from CW illumination.

## 4.2 Comparison of SSS and MM Theory with Detectibilities: SR versus SU

Having solved the simple square system for a single mode in $x$ ($M = 1$) with (18) and (25), mathematical induction[27] implies that there can be a closed form solution for $M > 1$ half-modes in x. Starting with (17) for a slowly sine swept system (SSS) and then adding only one mode along the short edge ($M = 1$), the CW integrated radiant flux return of (18) has a dependence on all the modes ($m = M = 1$; $n = 1, 2, \ldots N$). Expanding $M > 1$ past the $M = 1$ estimate (40) provides similar MM mode sensitivity for multi-modal systems compared to (25), as implied by Figure 10(b) where for $M > 1$ the values of $\Phi_e$ did increase above zero (for $M = 1$) until noticeable (○), as explained below.

$M = N$ results shown in green squares in the lower plot are all zero for $M = N = 1, 4, 9, \ldots$. The existence of structures with cut-off bandwidths above $M = N = 4$ becomes increasingly unlikely for higher numbers of perfectly square perfectly matched side modes with perfect physical and material properties (*super-symmetry*). In reality the modes associated with high bandwidth super-symmetry, which implies $M = N > 1$, should be detectibly imperfect and therefore different, even for a perfect, flat, square plate we attempt to constrain to $M = N = 1$, for any manufactured product no matter how perfectly machined. A possible exception would be if a prototype was specifically manufactured to exploit this effect using some kind of micrometer modified constraint system with actuator feedback, especially for 1-D vibration. The lower plot in Figure 10 shows three symmetry cases: $M = 0$ (1-D vibration) based on (42), $M = N$ (2-D super-symmetry) based on Equation 43, and $M \neq N$ (typical 2-D vibration) based on (44). The first two symmetry cases show up as flat lines in the plot in Figure 10, solid black dots ● and green squares □. The more likely distribution of modes, $M \neq N$, provided results shown in red circles ○ that vary for three or more modes by a small but measureable amount (see red dots ·, which are the variance $\sigma$ of the ○ values for $M \neq N$. To summarize and compare these cases the mode numbers $(m, n)$ and panel dimensions $(a, b)$ are collected into new mode numbers $(r, s)$ as coefficients of a generic spatial frequency $k_s$. These three expressions based on (17) (before SSS assumptions) and (18) produced this data for Figure 10:

For the 1-D case:
$$E_{\text{mod}}\big|_{M=0} \approx 2 \int_0^{10\mu m} \cos\left(\frac{4\pi}{\lambda} \sum_{r=1}^{N} \frac{\sin(rk_s y)}{\lambda}\right) dy \qquad (42)$$

2-D super-symmetry:
$$E_{\text{mod}}\big|_{M=N} \approx 2 \int_0^{10\mu m} \int_0^{10\mu m} \cos\left(\frac{4\pi}{\lambda} \sum_{r=1}^{M=N} \sum_{s=1}^{N} \frac{\sin(rk_s x)}{\lambda} \frac{\sin(rk_s y)}{\lambda}\right) dx dy \qquad (43)$$

2-D typical case:
$$E_{\text{mod}}\big|_{M \neq N} \approx 2 \int_0^{10\mu m} \int_0^{10\mu m} \cos\left(\frac{4\pi}{\lambda} \sum_{r=1}^{M} \sum_{s=1}^{N} \frac{\sin(rk_s x)}{\lambda} \frac{\sin(rk_s y)}{\lambda}\right) dx dy \qquad (44)$$

These calculations document structural and modal conditions that generate *spectral uniformity* (SU). The basic interference physics in Figure 11 shows SR and SE. Increasing the spot size of a nadir-directed probe causes a fluctuation in non-imaging return due to phase cancellation. SE occurs when the spot size exactly covers two anti-symmetric Chladni zone portions as in 'Sample 2.' For Sample 2, $A_{\text{spot}}/A_{\text{bar}}$ of 50% represents illumination of a full structural wavelength for a given mode ($N = 2$) resulting in SE. Partial illumination of a Chladni zone generates SR as in 'Sample 3.'

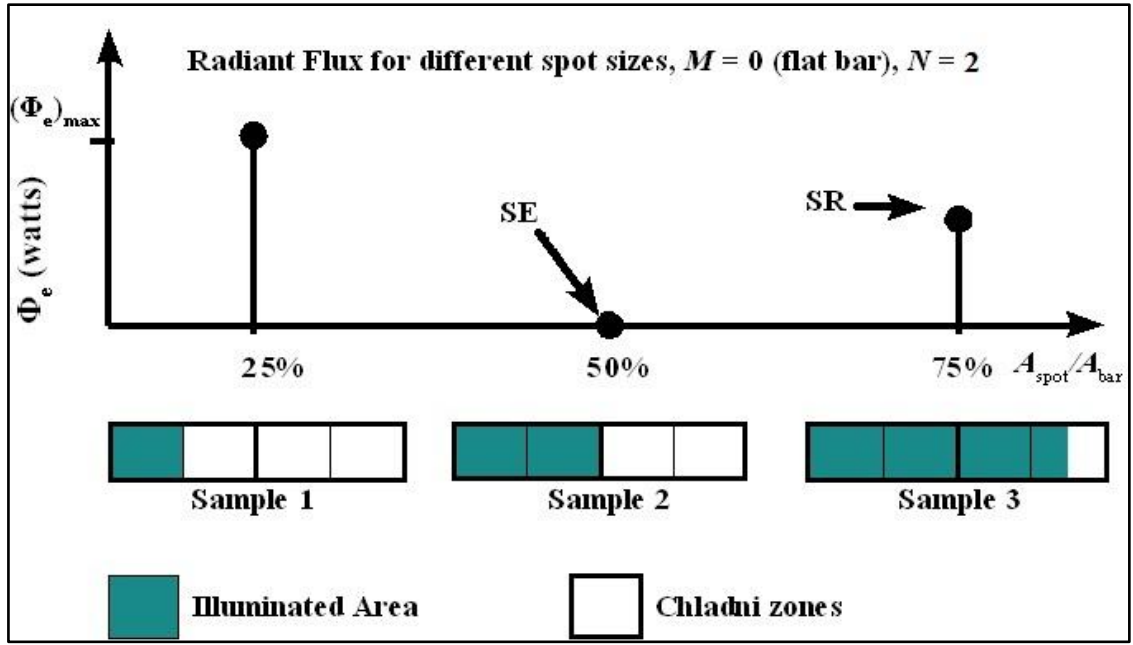

Figure 11. Spectral reduction and elimination as a function of spot size variation

Revisiting Table 2. Forms of spectral ID and classification deficiencies on page 5 and considering Figure 11 and Figure 12 a multi-dimensional sequence of ID and classification deficiency becomes apparent. By mathematical induction[27] we have just shown SR, SE and SU are negligible in reality except for super-symmetric modes ($M < 2$). This conclusion is further clarified by (44) considering the discussion below. The crosses in Figure 10(b) represent different mode combinations selected from $N = 1$ to $N = M − 1$ for all $M$ (the number of $x$-modes) arranged in the order of the total number of modes $N$. The slight variation in cross location ($\Delta \Phi_e$) provides a change in flux sufficient for identification. Even within the three tight groups of values, within the groups the values are shown to vary with the total number of modes. The red dots · scaled for clarity in Figure 10(b) are the variance $\sigma$ of the ○ values for $M \neq N$.

Some 2-D modes (both $M, N > 0$) still show SU as we might expect: $M = N = 1$ using (18) on page 25 for $M = 1$. Application of $N = 1$ causes the argument to be a single sine squared term. In the imaginary exponent the term $\sin^2\theta = ½ + \sin(\theta/2)$ provides a phase term factor $4\pi\varphi_n/2\lambda$ and a doubled spatial frequency modulation within which another Bessel function with a constant argument develops. So the SSS approximation validates the MM small deflection $M = N = 1$ case in that it has spectral uniformity with the green box □ at $M = 1$ for the MM radiant flux in Figure 10(b). The calculations are complicated but the concept is simple enough: compare to $N_{max} = 2$ in Figure 12. Broader bandwidth super-symmetry ($M = N > 1$) produces SU shown with □ at $N = 3$ and $N = 9$ in Figure 10(b), but the formulation is no longer as simple as $M = 1$ and the likelihood of $M = N$ super-symmetry drops to nil with growth in the number of modes (increasing bandwidth). Equations (42) and (43) above describe these cases. Instead of varying the illumination for one vibrating state as in Figure 11, Figure 12 shows an increase in the frequency while maintaining a full coverage illumination constant.

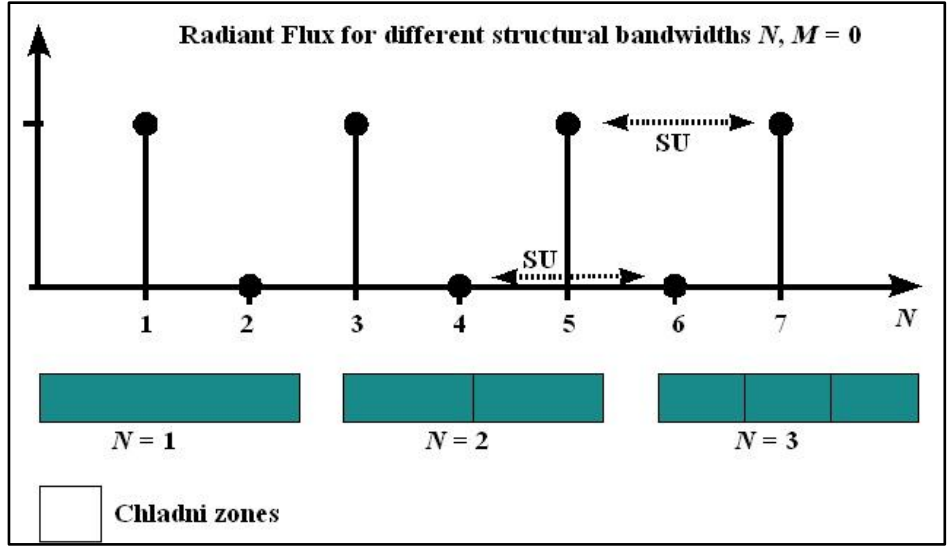

Figure 12. Structural vibration bandwidth *N* affects spectral uniformity, variation of Chladni zone size

For $M \neq N$ the ratio of possible configurations of maximum mode number to all configurations is $N(N-1)/N^2$. For example, with $N = 10$, the percentage of super-symmetric modes is $N/N^2 = 10\%$ of all perfect symmetry rectangles. These super-symmetric shapes can be more difficult to classify. And perfectly square flat panels are a negligible proportion of flat panels likely to be sensed. Further, flat panels themselves are a small portion of panels to be sensed. *The percentage of 'perfect types' of structural object features such as 'square,' 'flat,' and 'straight edges,' leads to a product of probabilities providing a crude estimate of overall classification error probability P of an optical sensor due to symmetry and super-symmetry:*

$$P_{\text{opticErr}}\begin{pmatrix} \text{perfect} \\ \text{plate} \end{pmatrix} = \frac{1}{N_{M=N}} \cdot \frac{1}{N_{\text{square}}} \cdot \frac{1}{N_{\text{flat}}} \cdot \frac{1}{N_{\text{straight}}} << 0.1 \qquad (45)$$

## 5.0 Discussion

Equation (45) reasonably estimates that the likelihood of super-symmetry in sensed panel vibrometry is negligible. In this case, when 1-D modes are unlikely, the structural elimination effect is so limited that to use it as a reason to avoid laser vibrometry proposals would be impractical.

Both SSS and MM show SU forms of phase averaging. In this regard they are similar to the spectral elimination created in the lab. The SE of the 1000 Hz line that Fl. Lt. Pepela created in the lab[2] uses an increase in spot size (Figure 2) and thus encompasses more return at different locations on the bar. So the larger spot size return undergoes a different combination of phase modulation than the point or pencil beam sensing return (Figure 1). By mathematical induction[27] the inclusion of return from a larger area of the bar provides even more overlap of particular phase values (many different deflection values) with which to modulate the probe beam.

The use of full coverage (large spot size) allows for reduced reliance on laser pointing and region hopping systems (tracking), and therefore, the potential for less expensive vehicle identification systems, based on laser vibrometry with simpler tracking systems.

One could conceive of a system, such as an active noise cancellation system, which could remove modes; it is possible as a counter-countermeasure or stealth scheme to at a minimum create "camouflage" SR. However, most detectors have a larger bandwidth than active systems have the capability to adequately remove modes—the energy has to go somewhere, and we have access to those regions that the energy goes to as well. *The thesis of this report is that a laser vibrometry detection system that chooses to use spectral ID on commercially manufactured product vibrations will have negligible chance of failure to classify when the pertinent signature modes are sensed.* The exception appears to be the precision case described above where extraordinary effort is made in design and near-perfect manufacturing to intentionally frustrate the classifier, an incredibly expensive endeavor. The current-day non-existence of such counter-countermeasures or stealth structures indicates the utility of laser vibrometry for vehicle identification and classification.

SR, SE and SU are also related in several ways because uniform response frustrates some spectral classifiers. First, the Neaman-Pearson theorem[35,36] validates single degree of freedom thresholds for likelihood vectors of independent identically distributed (iid) variable components. Next, a coupling of two separate features for each mode provides a 2-D classification system using a joint probability distribution that can be characterized by a Bhattacharya function for each resonant frequency. Two such features are response magnitude and center frequency location. Finally, the collection of at least a few modes for a vehicle vibration signature will layer an M-ary system on top of the joint distribution for each mode. SU acts to remove the joint distribution for each mode. The set of modes is sufficient for classification, but the detection power (probability of detection) can have been reduced dramatically for particular classifier that relied on the 2-D aspect per mode for adequate classification power.

To help determine possible degradation of vehicle identification capability using laser vibrometry, consider the effect of SR, SE and SU on system estimators.[35,36] If there are unavoidable natural effects from large spot size illumination for non-imaging detection, alternate detection systems might be necessary. Ngoya Pepela[2] produced a laboratory result of removing a single mode with a small increase in spot size from approximately 2% to 4% of the vibrating metal band area, a metal strip vibrating with structural 'bar' vibration modes (1-D modes). However, the work in this report shows that such SR is not an issue in the field; only for the super-symmetries described in the report does SR reduce the detection power.

On page 11, a paragraph described how issues with spatial coherence related to small spot size 'high frequency' laser sensing were unlikely to be an issue for this large spot size method of spectral ID operating on 'low frequency' vibration in the 50-250 Hz range. It would be useful to propose a measurement to verify these partially validated conclusions.

## 6.0 Conclusions

The main conclusion that the MM and SSS results shield identification and classification capability from ID and classification deficiency due to SR, SE and SU relies on several concepts. First, that deficiencies inherent in radiant return from super-symmetric structures are impractical to estimate because commercial products could not afford to build mathematically pure structures, they can only be "built" in computer models. Second, since very small spot sizes are immune from SR, SE and SU even for super-symmetric structures the article conservatively assumes a full illumination of the target. Further, the *observation* of SR, SE and SU, and *simulation* of same lead finally to the *calculation* of the electric field fluctuation for a particular instant in time due to structural modulation of the radiant return. The calculations show that modulated return is a function of the number of active modes, thus refuting SU and by implication SR and SE in the sensing of fully illuminated targets. The next sub-section elaborates on these underlying concepts to support the conclusion that commercial products are safely subject to spectral classification and identification.

## 6.1 Primary Conclusions

1. *Observation*: Radiant flux $\Phi_e$ in watts/cm$^2$ from a flat strip or bar has measured spectral reduction due to increased spot size; compare Figure 1 to Figure 2. The reason for this SR is similar by mathematical induction to spectral uniformity via the complete coverage (large spot size) non-imaged signal calculated by this plane wave theory. See the 'Discussion' section preceding this Conclusion section. The former is a variation in spot size (Figure 11) and the latter is a variation of modal number for a spot size that covers the entire target (Figure 12).

2. *Simulation*: For perfect rectangular plates fully illuminated by a large spot size, the higher frequency anti-symmetric and unsymmetrical modes were shown by simulation to have some spectral reduction similar to the phase averaging results of SSS and MM.[3] These results are time-based frequency response functions (FRF's) of the phase-modulated non-imaging return from simulation of an M1A1 tank armor plate that clatters microscopically on the hull. The vibration simulation of a nonlinear contact and transient finite element analysis (FEA) used NASTRAN$^{(TM)}$. The optical sensing simulation used Matlab$^{(TM)}$.

3. *Calculation*: $\Phi_e$ from a perfectly square plate simply supported on all edges has theoretical spectral reduction for super-symmetrical $M = N$ modes. This article's calculations show that spectral reduction for structures, that are not constrained to one-dimensional modes such as bars or metal strips, will require a precarious balance ('super-symmetry' in this document) where one dimension is constrained to a single mode ($M = 1$), and the structure has perfect edges, material, uniformity, and other expensive manufacture in order to attain spectral reduction. Except in slender bars ($M = 0$), SR is difficult to attain in perfect rectangular plates, and impractical to expect in manufactured (imperfect) vehicle skin. Figure 9 and Figure 10 support conclusions related to calculations in Table 3 below which is a short summary of the results.

Table 3. Modulation term, a function of bandwidth $N$ using different calculations (SSS vs. MM)

| Estimate type | Dimensionality | Equation | Received Field Fluctuation, $E_{mod}$ (Equation 46) |
|---|---|---|---|
| SSS | $M = 0$ | 17 (p. 25) | $E_{mod} = E_{in}(\pi (\Delta z)_{max} / N\lambda)^2 = E_{in} \zeta^2/N$ |
| SSS | $M = 1$ | 18 (p. 25) | $E_{mod} = E_{in}(\pi (\Delta z)_{max} / N\lambda)^2/2 = E_{in} \zeta^2/2N$ |
| MM | $M = 0$ | 24 (p. 29,25) | $E_{mod} = E_{in}(\pi (\Delta z)_{max} / N\lambda)^2 = E_{in} \zeta^2/N$ |
| MM | $M = 1$ | 25 (p. 30,25) | $E_{mod} = E_{in}(\pi (\Delta z)_{max} / N\lambda)^2/2 = E_{in} \zeta^2/2N$ |

Considering how the MM results (17) and (18) match the SSS results (24) and (25) even though the calculations are entirely different, the second order estimates for SU appear to be valid. These provide a numerical basis which, using Figure 11 and Figure 12, describe the physics behind SR and SE. The extreme measures required to manufacture a super-symmetric structure, except for bars, indicates that using vibrating surfaces for vehicle and vibration spectral ID if feasible. The relationships simplify using a dimensionless length $\zeta$:

$$\zeta \equiv \pi \frac{(\Delta z)_{max}}{\lambda} \quad \Rightarrow \quad E_{M=0} \approx E_{in} - E_{mod} = E_{in} - E_{in}\left(\frac{\zeta^2}{N}\right) \quad E_{M=1} \approx E_{in} - E_{in}\left(\frac{\zeta^2}{2N}\right)$$
(46)

4. *Conclusion*: **To the extent that modes do not have uniform modal participation (MPF's vary: $\varphi_m \neq \varphi_n$ for some $m \neq n$), the SSS and MM spectral reduction and elimination effects expected by theory and simulation bound by all the assumptions are practically removed.** Spectral "fingerprints" are even more unique than presented under UMP (see Figure 7).[22,23] Confluence of perfect and precarious conditions that would create a phase of return from different target structural vibration modes such that modal response no longer varied for full coverage (large spot size), or even for partial insensitivity of modes, is unlikely to be a natural modulation effect. From the MM section starting on page 31, extending to $M > 1$ (retaining $N > 1$) the results are far more complicated: $M, N > 1$ solutions do not provide simple monotonic results such as the correction term proportional to $1/N$ in (40). The mathematical reason was that integrals of all orders and all combinations do not allow cancellation in the sum as was used in the transition from (28) to (36) to (40). This complicated relationship due to the lack of cancellation aids spectral classification, even for perfectly rectangular plates as long as $M, N > 1$. The case for $M < 2$ is severely restrictive, intractable to produce even in a laboratory, except with typical simple 1-D bars where $M = 0$ (slender bars or stiff strips). Apart from the effects of $M =$ 0, which might be seen in the meager cross-section return from windshield wipers or whip antenna vibrations, the SE and related SR and SU effects are so limited that the existence in a few super-symmetric cases is inadequate to doubt the utility of laser vibrometry as a remote sensing method. Super-symmetry is rare in vehicle skins. *Therefore, SR and SU in the spectral classification and identification of ordinary structures is unlikely to be a practical concern.*

### 6.2 Secondary Conclusions

When should SR, SE, and SU be a concern? It may be easy to produce 1-D surrogates. However, there may be other methods to exclude 1-D structures from consideration for vehicle spectral ID.

Extending to practical structures where M > 1 (where N > 1) the calculations were more complicated and did not provide simple monotonic results (such as the correction term proportional to 1/*N* from Equations listed in Table 3) since integrals of all orders and all combinations do not allow cancellation in the sum used in the transition from (28) to (36) to (40). **This allows for spectral classification, even for perfectly rectangular wide plates** partly because for *M* > 1, modulation term $\psi_p$ is more complicated, (36) no longer allows sums of squares to factor out via $\int_0^b \left(\sum_p^N \psi_p(ky)\right)^2 dy$.

The lack of SR and SU is a necessary but not sufficient condition for acceptance criteria of CW return (CSC's,[3] spectra, and MAC's[15]) to be adequate classifiers for actual rectangular plates. These calculations show vibration classifiers built on large spot size return are incomplete classifiers for metal strips (bars) due to SR, yet adequate for structural ID of even perfectly rectangular plates where necessarily M > 1.

A separate result as a consequence of the derivation of the SSS expressions is the definition of a phase-modulation index $\delta_{mod}$. For a laser frequency of $\omega = c/2\pi\lambda$ we define in (26), copied here as (47) the expression for $\delta_{mod}$ for spectral identification using laser vibrometry. There is a variation in phase over the target surface in (*x*, *y*) space as deflection varies for each mode (*m*, *n*), limited by *M* < 2, vibrating at a frequency and $\omega_{m,n} = 2\pi f_{m,n}$ as in Yariv and Yeh:[26] $E_{out} = A_o \cos[\omega t + \delta \sin(\omega_{m,n})]$. The deflection normal to the plate in the *z*-direction is $w_{m,n}(x, y)$. Dimensionless length $\zeta$ defined in (46) is proportional to the number of wavelengths of deflection in the vibrating panel.

$$\delta_{mod} = \frac{2\pi w_{m,n}(x,y)}{\lambda} = \frac{\pi (\Delta z)_{m,n}(x,y)}{\lambda} \xrightarrow{UMP} \delta_{mod} < \left(\frac{(\Delta z)_{max}}{\lambda}\right) \frac{\pi}{\beta N} = \frac{\zeta}{\beta N} \qquad \beta = \begin{cases} 1 & M = 0 \\ 2 & M = 1 \end{cases}$$
(47)

This modulation index is cast in terms of the first-order amplitude modulations of the electric field *E*. *Since the detectors sense irradiance* in watts, proportional to the square of the field, the *second order terms* omitted for SSS (25) and MM (28) *are even more negligible* than they appear in those Equations. This modulation index may be a more appropriate indicator than those higher order terms for *E* field imply.

**Appendix A: System Coherence for Detectability**

This Appendix elaborates on the bounds of the calculations and simulations.

Optical roughness of the target can change the received pixel intensity statistics. The distances within which reflections from a target remain coherent are distances that bound a coherence cell. Assuming a range of 4 kilometers for a transmitting aperture approximately the same size as the receiving aperture, the number of correlation cells, parameter *M not to be confused* with the bandwidth *M* in this report, is in single digits, approximately $1 < M < 4$. As further explained by Goodman,[37] we might expect beam divergence to be larger than target extent at range, which drives *M* to the minimum (unity). The difference can be substantial, so a calculation estimating *M* is reasonable. For a single cell the underlying negative binomial distribution is the equivalent to a Bose-Einstein distribution, but as the number of correlation cells *M* increases, the underlying distribution "approaches a Poisson distribution" which is the photon irradiance statistics model typically assumed for lasers. Calculating *M* for a laser aperture of diameter a with detector width *d* at a range *R* makes use of Goodman's normalized range, $R\lambda/\pi a d$ = (4km)·(10$\mu$m) / $\pi$(5mm)·(10mm) ≈ 640, which drives the number of correlation cells down to the minimum of one (Goodman's Figure 4, his page 1096).[37] One practical result of such a low *M* is the intensity variance calculations for maximum a priori (MAP) estimators will use darker pixels than for a system with *M* larger than unity.[38]

## Appendix B: Adequacy of a Non-Unity Diagonal

This appendix refers only to a physical model as used by the simulation, item 3 in Table 1.

Figure 5 on page 9 shows that the diagonal of the cross-spectral covariance CSC($f_i$, $f_j$), the variance terms $\sigma_{i,i}^2(f)$, has deviations from unity that are small. Approximately nine 1.0 Hz width spikes that deviate from unity by more than 20% are within the first 100 Hz of this line plot of $S_{opt}(f) = \sigma_{i,i}^2(f) \otimes S_{vib}(f_v)$ in the Figure. Three of these spikes are more than double the average unity response. Conservatively estimate that 9%−3% = 6% of the spectrum undergoes at least 20% distortion and 3% undergoes at least a double response. A crude overall ensemble estimate of classification error for a single mode would be approximately (0.2 × 6/100) + (3 × 3/100) ≈ 10%. This result is conservative due to the nonlinearity in this model related to contact vibration which drives a 'contact' phase averaging classification error of approximately 10% related to both missed detection and false alarm. However, each additional mode used (incrementing cutoff modes $M$ and $N$ higher) provides a detection-fusion-related improvement in vehicle identification. An estimate of this reduced error of multi-mode classification for the use of three modes would be $0.1020^3 = 0.0011 = 0.11\%$. Therefore the diagonal of the CSC, $\sigma_{i,i}^2(f)$, is sufficiently close to unity that spectral identification should be possible with some attention paid to the small but unlikely error spikes in narrow spectral widths (that not more than one spike over-laps an important structural modal frequency). Otherwise the number of modes to use for ID would have to increase, which is not much of a problem.

A more numerical example:

Assume the chance of a mode occuring within a particular 2 Hz frequency bin in the 0-350 Hz range of interest to be *independent and identically distributed* (iid). Estimate that the spikes exceeding 40% from unity in Figure 5 on page 7 have the same width and number 40 within the 175 total frequency bins. Allowing zero wrong modes for a total of three modes used to identify (spectrally "fingerprint") the target, the chance of miss-classification is $40^3/(175 \cdot 174 \cdot 173) = 1.22\%$. Allowing one wrong mode out of four used for spectral ID the mis-classification chance is $40^3 \cdot 39/(175 \cdot 174 \cdot 173 \cdot 172) = 0.28\%$. Having discussed the reasonable cases, consider that with only two modes for identification the mis-classification probability is $40^2/(175 \cdot 174) = 5.3\%$, near the limit of tight 'engineering error.' Physical system characteristics that would violate iid include mode coupling and excitation of harmonics, which do occur. However, the mis-classification probabilities given above are reasonable given experience in automotive vibration design and testing, and they conform to the laser vibrometer spectral ID for the NATO study.[4]

As discussed on page 8, the simulation shown in Figure 4 resulted in a catalogue of results[3] that clearly show the modal features of the structure survive in the non-imaging return at the sensor for large spot probing of the M1A1 armor plate as a vehicle ID feature for IFF. This is a non-exhaustive example showing SR related to non-unity CSC values does not corrupt classification of this vibrating plate. Rarely are structures that are targets for ID as symmetrical as this rectangular armor plate. Therefore, while this argument cannot directly prove the negative (that there is no classification problem for any manufactured structure), by mathematical induction logic[27] the likelihood that more complicated structures would be harder to ID using spectral signatures is negligible.

**Acronyms, Abbreviations and Symbols**

AFIT   Air Force Institute of Technology of Air University, USAF, WPAFB, Dayton, Ohio
AFB    Air Force base
BC     Boundary condition
CSM   Closely spaced mode approximation
CW    Continuous wave laser illumination
$\mathcal{F}[\cdot]$   The Fourier transform operator
FE     Finite element, as in FE model
FEA    Finite element analysis (caution: much of the technology is over-marketed)
FRF    Frequency response function
ID     Identification
IFF     Identify Friend or Foe
iid     An independent and identically distributed statistical distribution or density
MAC   Modal acceptance criteria
MAP   Maximum a priori estimator
MM    Multi-modal estimate of the FRF, which for small deflection is far superior than SS
MPF   Mode participation function (MPF for the $n^{th}$ mode: $\Delta\varphi_n$)
NATO  North Atlantic treaty organization
$\mathcal{Re}[\cdot]$  The operator returning the real part of a complex number or function: $\mathcal{Re}[e^{j\theta}] = \cos\theta$
RHS   Right hand side (of an Equation), opposite of LHS
SE     Spectral elimination
SR     Spectral reduction
SSS    Slowly Swept Sine estimate for an incomplete form of the FRF
SU     Spectral uniformity
UMP   Uniform modal participation where all MPF's are equal to each other
USE    Uniform spectral elimination
WP     Wright-Patterson (AFB)

$\delta_{mod}$   Phase modulation index
$\varphi_r$     Relative phase of reflection
$\varphi$     Generic reflection phase as a function of distances from the source and to the sensor
$\lambda$     Wavelength, both structural wavelengths (approx. $1 - 20$ cm) and optical ($0.1 - 10\ \mu$m)
$\theta$     Generic phase, in Euler's representation $e^{j\theta}$, within which are sums of vibration modes
$\sigma$     The variance of radiant flux (in Figure 10)
$\omega$     Angular speed, $\omega = 2\pi f$, often called the 'frequency' but strictly speaking is a speed, $2\pi$ radians per cycle
$\psi$     Generic function, $\psi = \Delta\varphi_n \sin kx$ for example, to show trigonometric identities
$\zeta$     A dimensionless length which simplifies the main result of this work (Table 3, p. 44)

$\Delta\varphi_n$   MPF for mode $n$. $\Delta\varphi_n \equiv \Delta z_{max}/2N$
$\Delta z_{max}$  Full range of $z$ motion of a point on the target surface, vibration normal to the surface
$\Phi_e$     Radiant flux in watts, the spatial integration of irradiance or exitance

$a, b$   Dimensions (values of width and length) of a structural plate or panel
$h[n]$   Impulse response of a [discrete] variable
$k$     Propagation constant, $k = 2\pi/\lambda$, this spatial frequency is a *wavenumber* if units are cm$^{-1}$.

| | |
|---|---|
| $m$ | Mode number for vibration along $x$ ($\sin kx$) |
| $n$ | Mode number for vibration along $y$ ($\sin ky$) |
| $w$ | Vibration displacement in the $z$-direction, amplitude for variation $\sin \omega t \cdot \sin kx \cdot \sin ky$ |
| | |
| cm | centimeters |
| cz | Subscript to represent the Chladni zone (reference footnote ** on page 20) |
| km | kilometers |
| m | meters |
| mm | millimeters |
| ms | milliseconds |
| s | seconds |
| | |
| $E_e$ | Irradiance (watts/cm$^2$), power impinging on a sensor or object, related to $E$ field squared |
| $E_{in}$ | Electric field incoming $E_{in}$ is related to $E_e$, modulated $E_{mod}$ |
| $E_m$ | Electric field $E_{in}$ is related to $E_e$, modulated $E_{mod}$ is related to $M_e$, or other values |
| $E_{mod}$ | Modulated electric field W/m – power $M_e$ or $E_e$ in watts (without subscript e) |
| $H[f_i]$ | Transfer function of a [discrete] variable, the Fourier dual of the impulse response $h$ |
| $M$ | Maximum mode number for vibration along $x$, max $m = M$, the $x$-direction cutoff |
| $M$ | Number of correlation cells that the receiving aperture contains. Appendix A. |
| $M_e$ | Exitance (watts/cm$^2$), power exiting, radiating (usually) or reflecting from an object |
| $N$ | Maximum mode number for vibration along $x$, max $n = N$, the $y$-direction cutoff |
| $S_{opt}(f)$ | Power spectral density for optical return as a function of frequency, watts per Hz |
| $S_{vib}(f)$ | Power spectral density for structural as a function of frequency, g$^2$ per Hz (g = 9.81 m/s$^2$) |
| | |
| 1-D | One dimensional |
| 2-D | Two dimensional |

**References**

1. Saleh, Bahaa E.A. and Teich, Malvin Carl. *Fundamentals of Photonics* Wiley Inter-Science, John Wiley & Sons, 111 River Street, Hoboken, NJ 07030, ISBN 978-0-471-34832-9, $2^{nd}$ edition, 2007. [Cited herein from pages 1, 2,10.]

2. Pepela, Ngoya, "Effect of Multi-Mode Vibration on Signature Estimation Using a Laser Vibration Sensor", Master's thesis, Air Force Institute of Technology ENY, Wright Patterson AFB OH 45433 – 7321, December 2003. AFITGEENP03-02, p. 4-2 ff, p. 6-1. [Cited herein from pages 1, 2, 3, 4, 5]

3. Kobold, M.C., "Laser Covariance Vibrometry For Unsymmetrical Mode Detection". Electrical Engineering Master's Thesis(Optical Control and Signal Processing), Air Force Institute of Technology, Wright Patterson AFB OH 45433-7321, AFIT/GE/ENG/06-61, September 2006. p. 31, 66, 71, 73, 74, 119 (fig. 25), 157 (SR). This 2006 thesis at the Air Force Institute of Technology (AFIT) analyzed return from increased ranges using a finite element analysis (FEA) to simulate the vibration surface and a Matlab system to modulate and propagate the laser return to a simulated sensor. The presentation for this thesis was given as "Laser Covariance Vibrometry for symmetrical mode detection" to an internal AFRL Sensors Directorate.

4. Olsson, Andreas, *Target Recognition By Vibrometry With A Coherent Laser Radar*. Examensarbete, Linköpings University, Institutionen för Systemteknik, 581 83 LINKÖPING, 04 2003. www.ep.liu.se/exjobb/isy/2003/3050. [Cited herein from pages 4, 6, 7, 14, 15, 23] "The distance to the target varied between 530 and 3200 meters." Page 44. [Cited herein from pages 2, 6, 7, 11, 14, 16, 25]

5. Pepela, N., Cobb, R., and Marciniak, M., Polytech OFV-3001 Laser Doppler Vibration Sensor System. used in the 2002 AFIT M.S. thesis of RAAF Flight Lt. Pepela. [Cited herein from pages 2, 3, 16]

6. Armstrong, E., *Multifunction Ladar Integration And Demonstration System*. USAF/AFRL/SN, WPAFB, OH, hardware used in Dierking report, 7 Dec. 2003. [Cited herein from pages 2, 7]

7. Dierking, Matthew; Muse, Robert; Barnes, Lawrence; Seal, Michael, and Armstrong, E., *AFRL Vibrometry Assessment Outline*. Contract AFRL-SN-WP-TR-2003-XXXX (now AFRL-RY…), USAF/AFRL/SN Combat Identification Technology Branch, 3109 P Street, Building 622, WPAFB, OH 45433–7700, Dec 2003. [Cited herein from page 2]

8. Pepela, N. and Marciniak, M., E-mail response with permission to publish [Wed 27 June 2007,10 11 am]. Figures 4.5 and 4.6 in RAAF Flight Lt. Pepela AFIT M.S. thesis, June 2007.5 [Cited herein from pages 2, 4]

9.   Zoiros, Kyriakos E., O'Riordan, Colm, and Connelly, Michael J., "Semiconductor Optical Amplifier Pattern Effect Suppression Using Birefringent Fiber Loop". *IEEE Photonics Technology Letters*, 22(4):221–223, February 2010. [Cited herein from page 5]

10.   Bowman, John C. B., Shadwick, A., and Morrison, P.J., "Spectral Reduction: A Statistical Description of Turbulence". Microsoft Technical Reports, Dec 1999. [Cited herein from page 5]

11.   Thomson, William Tyrrell, *Theory of Vibration With Applications*, Prentice Hall, Englewood Cliffs, NJ 07632, 1988. PSD is on page 367. [Cited herein from page 9]

12.   Goodman, Joseph W., Statistical Optics. Wiley - Interscience Publications, John Wiley & Sons, Inc., New York, 1985. Page 80 discusses cross-spectral densities. [Cited herein from page 9]

13.   Kobold, Michael C. "Videst.m Based on NATO Data in the 2003 Swedish Dissertation The Andreas Olsson".[4] Unclassified Matlab ROC code used to propose laser vibrometry to DARPA, Videst.m, Dec 2005. [Cited herein from pages 14 and 16]

14.  Allemang, R.J., *Investigation of Some Multiple Input/Output Frequency Response Function Experimental Modal Analysis Techniques*. PhD thesis, The University of Cincinnati, Structural Dynamics Research Lab., Mech., Ind., and Nuclear Eng., U. of Cincinnati, OH 45221–0072, 1980. [Cited herein from page 10]

15.   Allemang, R.J..,The Modal Assurance Criterion Twenty Years of Use And Abuse. Sound and Vibration, 37(9):8–17, Aug 2003. [Cited herein from pages 11, 33]

16.   Allemang, Randall J., "Chapter 21, Modal Analysis and Testing". McGraw – Hill, New York, 4th edition, 1996. Cyril M. Harris, editor, *Shock and Vibration Handbook*. [Cited herein from pages 11, 15, 16, 19]

17.   Curtis, Allen J. and Lust, Steven D., "Chapter 20, Concepts in Vibration Data Analysis". McGraw-Hill, New York, 4th edition, 1996. Cyril M. Harris, editor, *Shock and Vibration Handbook*. [Cited herein from page 11]

18.   Rubin, Sheldon. "Chapter 23, Concepts in Shock Data Analysis". McGraw – Hill, New York, 4th edition, 1996. Cyril M. Harris, editor, *Shock and Vibration Handbook* [Cited herein from pages 11]

19.   Allemang, Randall J., Structural Dynamics Research Laboratory, University of Cincinnati. Personal Correspondence, Cincinnati, OH 45221-0012, www .sdrl .uc .edu, 2006.[Cited herein from page 11, 13]

20.   Kobold, M.C., "Shear Deflection Estimate Of Soil Vibration Above A Buried Object", NSWC PCD Technical Note TN-13-002, Submitted March 2013. [Cited herein from pages 11 and 16]

21.     Goodman, Joseph W. *Introduction to Fourier Optics*. McGraw–Hill Book Company, San Fransisco, 1968.  Parseval's, the convolution, and the autocorrelation theories are described on page 10. [Cited herein from page 14]

22.     Brughmans, ir M., Lembregts, ir R., and Furini, ir F., "Modal Test on the Pininfarina Concept Car Body 'Ethos 1'," Part of the development of a nonlinear frequency response program for simulating vehicle ride comfort (pininfarina). *1995 MSC World Users Conference Proceedings*, paper 5. MacNeal-Schwendler Corporation, www .mscsoftware .com (remove spaces to activate link). [Cited herein from page 16, 25, 44]

23.     Lammens, Stephan; Brughmans Marc; Leuridan, Jans and Sas, Paul, "Application of a FRF based Model Updating Technique for the Validation of Finite Element Models of Components of the Automotive Industry,"  Part of the development of a nonlinear frequency response program for simulating vehicle ride comfort (pininfarina).  *1995 MSC World Users Conference Proceedings*, Costa Mesa, CA, 1995, paper 7. MacNeal-Schwendler Corporation, www .mscsoftware .com (remove spaces to activate link). [Cited herein from page 16, 25, 44]

24.     Kobold, M.C., "Scintillation Response Sensitivities; Effects of Turbulence on Laser Sensing of Soil Surface Vibration," NSWC PCD Technical Report, TR-2013/012, Submitted September 2013.  Vibration modes are on page 4 especially in Figure 4. [Cited herein from pages 16]

25.     Strutt, John William, the Baron Rayleigh Sc.D. F.R.S., *The Theory of Sound, Volume I*, originally published in 1877, second revised edition, Dover Publications, NY, Printed by General Publishing Co. Ltd., 30 Lesmill Road, Don Mills, Toronto, Ontario, 1945.  The concept of modes is on page 107 and stationary states on page 110.  Concepts leading to Chladni zones on page 345-349, citing Chladni on page 367 and tying into superposition on page 377 – the footnote § herein on page 20 may be particularly instructive.  Page numbers provided because the text has no index. [Cited herein from pages 12, 16, 18, 20]

26.     Yariv, Amnon, and Yeh, Poci. *Optical Waves in Crystals, Propagation and Control of Laser Radiation*. Wiley – Interscience, a John Wiley & Sons, Inc. Publication, New York, 2003. p. 243 ff. [Cited herein from pages 16, 24, 31]

27.     James, Glen and Robert, *Mathematics Dictionary*.  Fourth edition, Van Nostrand Reinhold, New York,ISBN 0-442-24091-0, 1976.  Fubini is on page 159 showing how to compute a double integral from iterated integrals which implies rules for interchange of the order of integration. Mathematical induction is on page 196.   [Cited herein from page 18, 25-28, 34, 39, 40, 42 and B-1]

28.     Svelto, Orazio, *Principles of Lasers*. 4th edition, Plenum Press, 233 Spring Street, New York, NY 10013, www dot plenum dot com, 1998. Control of Finesse, F, page 143. The mode-lock E field sum of modulation providing a sum of modes is (8.6.2) on page 332. The basic phase relation $\varphi n - \varphi n-1 = \varphi$ (8.6.1) is a uniform mode spacing assumption, pages 331-335. [Cited herein from pages 22, 28].

29.   Arfken, George. *Mathematical Methods for Physicists*. Academic Press, Inc., Orlando, FL, 32887, 2nd edition, 1985. [Cited herein from page 22]

30.   Abramowitz, Milton, and Stegun, Irene A., "Handbook of Mathematical Functions". National Bureau of Standards, U. S. Government Printing Office, Washington D. C. 20402, 1964. p. 42, p.360 # 9.1.20, p. 534 #4. [Cited herein from pages 22, 23, 27]

31.   Young, Warren C., *Roark's Formulas for Stress & Strain*, 6th Ed., ISBN 0-07-072541-1, McGraw-Hill, Inc., TA 407.2.R6, 1989. See Glossary for 'fixity.' Pages 62 (# 2.), 100 (# 1.), and 715 (# 3.) versus 717 (#'s 15-17).   [Cited herein from pages 30, 33, 34]

32.   Thomson, William Tyrrell. *Theory of Vibration With Applications*. Prentice Hall, Englewood Cliffs, NJ 07632, 1988.  Page 223, Figure 8.4-2.   [Cited herein from page 30]

33.   Proakis, John G., and Manolakis, Dimitris I., *Digital Signal Processing: Principles, Algorithms, and Applications*, Prentice-Hall, Inc. (Simon & Schuster), Upper Saddle River, New Jersey, 07458, 1996.  Page 414.   [Cited herein from page 32]

34.   Rubin, Sheldon, "CRC Standard Mathematical Tables," Editor Samuel M. Selby, 21st edition, The Chemical Rubber Co., 18901 Cranwood Parkway, Cleveland, Ohio 44128, 1973.  Page 437 (#316) and 462 (#629).   [Cited herein from page 33, 34]

35.   Scharf, Louis L., *Statistical Signal Processing*, Addison -- Wesley Publishing Company, Inc., Reading, MA, 1991. UMP is at page 124.   [Cited herein from page 42, 43]

36.   vanTrees, Harry L., *Detection, Estimation, and Modulation Theory, Part I (Volume I of V): Detection, Estimation, and Linear Modulation Theory*, John Wiley & Sons, Inc., New York, 1968.  UMP is on page 89.   [Cited herein from page 42, 43]

37.   Goodman, Joseph W., "Some Effects of Target-Induced Scintillation on Optical Radar Performance." *Proceedings of the IEEE*, 22(4):1688–1700, November 1965.   [Cited herein from page A-1]

38.   MacDonald, Adam, Cain Stephen C., and Armstrong Ernest E., "Maximum A Posteriori Image Seeing Condition Estimation From Partially Coherent Two-Dimensional Light Detection And Ranging Images". *Optical Engineering*, 45(8):086201–1, –13, August 2006.   [Cited herein from page A-1]

39.   The standard FEA codes are driven by commercial companies, their central support groups, and industry coalitions both official and ad hoc.  This group includes: ABAQUS, ADINA, ALGOR, ANSYS, ASTROS, COSMOS/M, CSA/NASTRAN, FEMAP, GIFTS, I-DEAS, IMAGES, LS/DYNA, MARC, Mechanica (Rasna), MSC/NASTRAN, mTAB/SAP386, NISA, PATRAN STARDYNE, and UAI/NASTRAN, and European codes ASAS, ASKA, BERSAFE, CASTOR, DIANA, FAM, LUSAS, PAM-CRASH, PERMAS, SAMCEF, SESAM, SOLVIA, SYSTUS, and Boundary codes EASYBEA and LMS, and fluid FEA codes CENTRIC, FIDAP, FLOTRAN, PHLEX, and RAMPANT.  These particular features of FEA including modeling (fitting an arc through three points of geometry before meshing) and display of multiple mode shapes with their frequencies are not features available with COMSOL 4.3.  It does have modules that communicate with Matlab (most of the FEA codes do not) and it has some nice modules, but it does not integrate the package into something a production engineering group can use on drawings in the way the standard FEA codes can.  Additionally, COMSOL requires the user create internal logic to define the eigenfrequency normalization because total mass is not calculated automatically (something FEA codes have done automatically since before 1985).

40. M H Schneider and R J Feldes and J R Halcomb and C C Hoff, "Stability analysis of perfect and imperfect cylinders using MSC/NASTRAN linear and nonlinear buckling, paper 27, MSC "World" Users' Conference Proceedings, 1995. Table 1 shows that the ratio of perfect (typical FEA) to imperfect (more like reality) cylinder buckling load results using nonlinear buckling (linear has too much error) for two cases: Experiment (2.55 & 1.84) and FEA (2.53 & 1.81). Adobe copy of cylindersHaveImperfectionSimulationMissesSchneiderNASTRAN95.pdf available upon request.  [Cited herein from page 7]

41. Sir Edward R. Henry C.S.I., "Classification and Uses of Finger Prints," London: George Rutledge & Sons, Ltd., 1900.  Part II describes the classification system and page 62 describes the mathematics of classification probabilities of that era with results of proven cases on page 68. [Cited herein from page 11]

42. Max Born and Emil Wolf. "Principles of optics; Electromagnetic theory of propagation interference and diffraction of light. Cambridge U. Press," The Edinburgh Building, Cambridge, CB2 2RU, UK, www.cup.cam.ac.uk, seventh (expanded) edition, 1999. Page 523 et.seq. discusses alternate vector base systems that are adequate but not orthonormal (Zernike modes) with a table on 530.